\documentclass[journal]{IEEEtran} 
\IEEEoverridecommandlockouts
\usepackage{cite}
\usepackage{amsmath,amssymb,amsfonts}
\usepackage{algorithmic}
\usepackage{algorithm}
\usepackage{graphicx}
\usepackage{textcomp}
\usepackage{xcolor}
\usepackage{pifont}
\usepackage{cases}
\usepackage{url}
\usepackage{multirow}
\usepackage{eqparbox}
\usepackage{subfigure}
\usepackage[switch,pagewise]{lineno}
\usepackage{geometry}
\begin{document}

\title{SEAL: A Strategy-Proof and Privacy-Preserving UAV Computation Offloading Framework}

\author{Yuntao~Wang, Zhou~Su, Tom H. Luan, Jiliang Li, Qichao Xu, and Ruidong Li
\thanks{Yuntao Wang, Zhou Su, Tom H. Luan, and Jiliang Li are with the School of Cyber Science and Engineering, Xi'an Jiaotong University, Xi'an, China}
\thanks{Qichao Xu is with the School of Mechatronic Engineering and Automation, Shanghai University, Shanghai, China}
\thanks{Ruidong Li is with the College of Science and Engineering, Kanazawa University, Kanazawa, Japan}
}

\maketitle

\begin{abstract}
Due to the limited battery and computing resource, offloading unmanned aerial vehicles (UAVs)' computation tasks to ground infrastructure, e.g., vehicles, is a fundamental framework. Under such an open and untrusted environment, vehicles are reluctant to share their computing resource unless provisioning strong incentives, privacy protection, and fairness guarantee. Precisely, without strategy-proofness guarantee, the strategic vehicles can overclaim participation costs so as to conduct market manipulation. Without the fairness provision, vehicles can deliberately abort the assigned tasks without any punishments, and UAVs can refuse to pay by the end, causing an exchange dilemma. Lastly, the strategy-proofness and fairness provision typically require transparent payment/task results exchange under public audit, which may disclose sensitive information of vehicles and make the privacy preservation a foremost issue. To achieve the three design goals, we propose {SEAL}, an integrated framework to address {\underline{\textbf{S}}trategy-proof, fair, and privacy-pr\underline{\textbf{E}}serving U\underline{\textbf{A}}V computation off\underline{\textbf{L}}oading}. SEAL deploys a strategy-proof reverse combinatorial auction mechanism to optimize UAVs' task offloading under practical constraints while ensuring economic-robustness and polynomial-time efficiency. Based on smart contracts and hashchain micropayment, SEAL implements a fair on-chain exchange protocol to realize the atomic completion of batch payments and computing results in multi-round auctions. In addition, a privacy-preserving off-chain auction protocol is devised with the assistance of the trusted processor to efficiently protect vehicles' bid privacy. Using rigorous theoretical analysis and extensive simulations, we validate that SEAL can effectively prevent vehicles from manipulating, ensure privacy protection and fairness, improve the offloading efficiency, and reduce UAV's energy costs and expenses with low overheads.
\end{abstract}


\IEEEpeerreviewmaketitle
\section{Introduction}
Recently, unmanned aerial vehicles (UAVs) are gaining growing interest in enabling various smart city applications such as traffic surveillance and disaster rescue \cite{10106022,8977328,9743346}.
Thanks to the high agility, low cost, and line-of-sight (LoS) transmissions, UAVs equipped with wealthy sensors can be flexibly dispatched for data collection and environment perception in areas which are inaccessible or hazardous for humans in an on-demand manner \cite{9625737}.
For example, in disaster rescue, UAVs usually follow preset flying routes to visit all disaster sites and perform several missions (e.g., survivor detection and target tracking) at each location \cite{9696188}.

\begin{figure}[!t]\setlength{\abovecaptionskip}{-0.05cm}
\centering
  \includegraphics[width=8.cm]{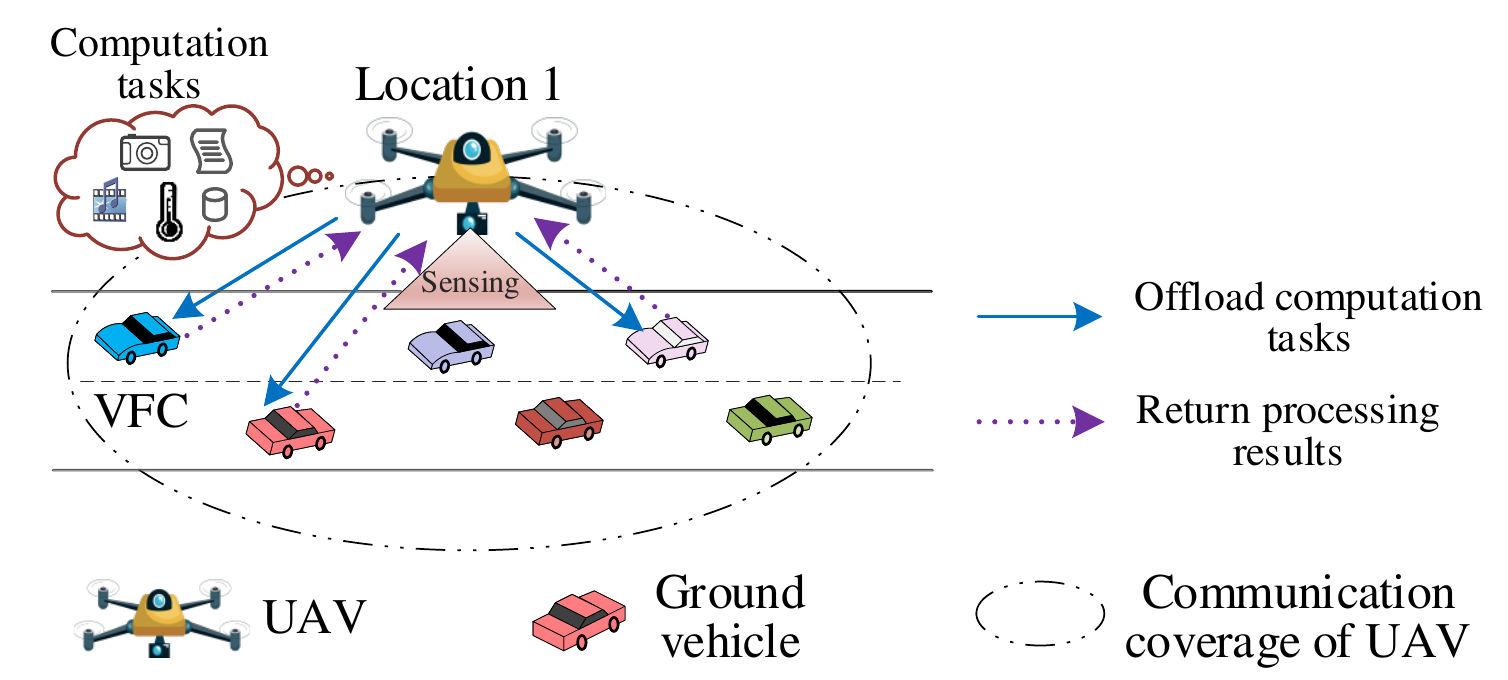}\\
  \caption{An illustrating example of UAV computation offloading based on vehicular fog computing (VFC).} \label{fig:systemmodel}\vspace{-0.45cm}
\end{figure}

Due to the size and weight limitations, UAV's battery capacity is usually constrained. For instance, the battery capacity of a small UAV with {a payload of} 300g is about 5200mAh, which supports a maximum flight endurance of 90 minutes \cite{9743346}.
Moreover, the real-time compute-intensive tasks such as image and video processing can greatly shorten UAVs' endurance time, thereby restricting their flying time and distance.
Thereby, efficient computation offloading is urgently needed for UAVs in performing persistent missions.
In {the} literature, UAV's computation missions are conventionally offloaded to the remote cloud for processing \cite{8981670,8723343}, and then the computing results are delivered back to the UAV.
Nevertheless, it incurs a long network delay in data transmission to/from the remote cloud, especially for latency-sensitive tasks such as traffic monitoring. To trade off the response latency and endurance time, a plausible solution is offloading UAVs' heavy computations to fog infrastructures \cite{9043589,8654698} such as roadside units and {Wi-Fi} access points. However, it highly depends on and bears the high deployment cost of additional ground infrastructures. 
Besides, the computing capacity of fog nodes is also limited to perform the heavy offloaded tasks before {the} expiration time.
Fortunately, vehicular fog computing (VFC) has been envisioned as a feasible and cost-effective solution by exploiting ground vehicles as moving fog nodes and offloading deadline-driven tasks to nearby vehicles with idle computation resources \cite{9696188,8977328}, as shown in Fig.~\ref{fig:systemmodel}.
Under the VFC paradigm, due to the proximity to end users, controllable vehicle mobility, and dense geographical distribution, it brings more convenience in offloading computation tasks produced by UAVs with reduced service delay.

To practically deploy VFC for collaborative computation offloading, incentive design is one of the fundamental issues.
As participating in such offloading tasks usually {consumes} considerable computation and battery resources, vehicles will be reluctant to share their computing resource without satisfactory incentives.
So far, various incentive mechanisms \cite{8660495,6732930,9292475} have been proposed to motivate users' cooperation in air-ground networking and lots of them are based on auction theory, in which the UAV is the service buyer and vehicles are service sellers.
However, selfish and strategic vehicles may misreport their participation costs and submit untruthful bids to claim more compensations so as to conduct market manipulation \cite{7559774}, thereby damping the enthusiasm of honest vehicles and raising the necessity of strategy-proof (or truthful) auctions.
In addition to strategy-proofness, fairness is another essential concern that may hinder vehicles' participation \cite{9296813}.
In practice, vehicles may deliberately withdraw assigned tasks to {seek high revenues. For example, a malicious vehicle with constrained computing resource may participate in multiple auctions for different UAVs, and it may abort the currently assigned task for a certain UAV and perform another assigned computing task for another UAV to gain higher benefits.}
Meanwhile, the ignorance of fairness in auction design may cause an exchange dilemma due to a lack of mutual trust, where neither the UAV nor vehicles are willing to initiate the payment/results transfer.
Besides, the strategy-proofness and fairness provision typically require transparent exchange of payment and task results under public audit, which may leak a great deal of vehicle users' sensitive information (including cost type, task preference, and resource capacity). Thereby, vehicle's competitiveness and resource strategies can be estimated by other rivals or adversaries, making the privacy preservation a foremost issue.

In the literature, existing privacy-preserving auction approaches \cite{7118253,8493354,8675981,9354536,8489932,8169049} are {mainly} based on the common assumption that participants will not abort the assigned tasks and there is no friction in exchanging the task results and rewards, which usually do not hold in realistic environments. The violation of fairness can cause the exchange dilemma (e.g., refuse to pay for computing results) and malicious dropout (e.g., abort assigned tasks), thereby causing loss to honest users and a task failure. Besides, existing mechanisms mainly offer single-parameter truthfulness (i.e., prevent strategic bid prices) for homogeneous (or identical) tasks/items. 
Under vehicle-assisted UAV computation offloading scenarios, the network environment can be time-varying, causing high heterogeneity and dynamics of tasks in terms of vehicular dwell time and offloading latency. In addition, UAVs usually require truthfulness for both computation resource supply and bid price of vehicles, resulting in a demand for combinatorial truthfulness.
Hence, it remains an open and vital issue to design a combinatorial strategy-proof computation offloading framework with fairness and privacy guarantees for UAVs under the VFC.

In this paper, we propose {SEAL}, an integrated framework to promote {\underline{\textbf{S}}trategy-proof, fair, and privacy-pr\underline{\textbf{E}}serving U\underline{\textbf{A}}V computation off\underline{\textbf{L}}oading} for general UAV applications. 
Specifically, we first present a VFC-based collaborative architecture to facilitate UAVs' on-demand computation offloading, and then investigate a single-minded reverse combinatorial (SRC) auction framework for computation task scheduling with combinatorial truthfulness guarantees of resource supply and bid price.
Afterward, to ensure fairness in the entire auction cycle, we resort to the smart contract technology and devise an on-chain fair exchange protocol to ensure both \emph{exchange fairness} (i.e., prevent exchange dilemma in the payment/task results delivery) and \emph{participation fairness} (i.e., prevent malicious dropout of assigned tasks) among distrustful parties.
When implementing SRC auctions in smart contract scripts, it requires vehicles' truthful bid input to be public for audit, which violates the privacy of vehicles. An off-chain auction execution mechanism is further developed in smart contract systems to effectively preserve vehicles' bid privacy with the assistance of the trusted processor equipped on the auctioneer (i.e., UAV). In addition, for efficient on-chain and off-chain orchestration, we devise a commit-then-claim mechanism with batch payment in smart contracts to ensure transactional atomicity and enhance trading efficiency under frequent micropayments.

To summarize, the main contributions are three-fold.
\begin{itemize}
  \item We propose SEAL to facilitate secure and efficient UAV computation offloading with two improvements: 1) a VFC-based SRC auction mechanism with high flexibility, on-demand deployment, and low response delay; and 2) an on-chain and off-chain cooperation mechanism to protect user privacy and enable fairness in the entire auction cycle with low system overheads.
  \item We consider the network dynamics due to the high mobility of UAVs and vehicles in the auction design to achieve both combinatorial strategy-proofness and near-optimal UAV cost minimization in practical offloading applications. We design a series of novel fair protocols based on smart contracts to forbid selfish bidders and ensure trust-free delivery of payments (to vehicles) and task results (to UAV).
      To further improve efficiency in multi-round SRC auctions, a commit-then-claim mechanism with hashchain-based batch payment is developed to reduce operational cost of smart contracts by sequentially delivering hash values (i.e., paywords) as micropayment commitments in an off-chain manner and claiming the due payment in an on-chain manner. 
  \item We theoretically analyze the property of SEAL and rigorously prove its capability in privacy protection, fairness, combinatorial strategy-proofness, and computation efficiency. We also conduct extensive simulations to verify the feasibility and effectiveness of SEAL. Numerical results show that SEAL can {reduce system overhead, alleviate UAV's cost, defend against strategic vehicles, and enhance offloading efficiency,} compared with conventional schemes.
\end{itemize}

The remainder of this paper is organized as follows. Section \ref{sec:RELATEDWORK} reviews the related work. Section \ref{sec:SYSTEMMODEL} introduces the system model. We elaborate the detailed design of SEAL and theoretically analyze its property in Section \ref{sec:FRAMEWORK}. We present the numerical results in Section \ref{sec:SIMULATION}. Section \ref{sec:CONSLUSION} closes the paper with conclusions.

\begin{table*}[!t]\setlength{\abovecaptionskip}{-0.0cm}
\caption{Existing truthful and privacy-preserving auction approaches: A comparative summary}\label{tableCMP}
\resizebox{1.01\linewidth}{!}{
\begin{tabular}{lcccccccccc}
\hline
\textbf{Ref.} & \textbf{\begin{tabular}[c]{@{}c@{}}Auction\\ Type\end{tabular}}                                             & \textbf{\begin{tabular}[c]{@{}c@{}}Bid Price\\ Truthfulness\end{tabular}} & \textbf{\begin{tabular}[c]{@{}c@{}}Combinatorial\\ Truthfulness\end{tabular}} & \textbf{\begin{tabular}[c]{@{}c@{}}Bid Privacy\\ Protection\end{tabular}} & \textbf{\begin{tabular}[c]{@{}c@{}}Participation\\ Fairness\end{tabular}} & \textbf{\begin{tabular}[c]{@{}c@{}}Exchange\\ Fairness\end{tabular}} & \textbf{\begin{tabular}[c]{@{}c@{}}Task/Item\\ Heterogeneity\end{tabular}} & \textbf{\begin{tabular}[c]{@{}c@{}}Comp.\&Comm.\\ Overhead\end{tabular}} & \textbf{\begin{tabular}[c]{@{}c@{}}Bid Utility\\ (Availability)\end{tabular}} & \textbf{Scenario}                                                          \\ \hline
PISA \cite{7118253}         & single                                                            & $\checkmark$                                                              & $\times$                                                                          & $\checkmark$                                                              & $\times$                                                                  & $\times$                                                             & $\times$                                                              & High                                                                     & High                                                                          & spectrum market                                                        \\ \hline
ARMOR \cite{8493354}        & combinatorial                                                     & $\checkmark$                                                              & $\times$                                                                          & $\checkmark$                                                              & $\times$                                                                  & $\times$                                                             & $\checkmark$                                                          & High                                                                     & High                                                                          & spectrum market                                                        \\ \hline
PS-TAHES \cite{8675981}     & double                                                            & $\checkmark$                                                              & $\times$                                                                          & $\checkmark$                                                              & $\times$                                                                  & $\times$                                                             & $\checkmark$                                                          & High                                                                     & High                                                                          & spectrum market                                                        \\ \hline
V2GEx \cite{9354536}        & ---                                                              & $\times$                                                                  & $\times$                                                                          & $\checkmark$                                                              & $\times$                                                                  & $\checkmark$                                                         & $\times$                                                              & High                                                                     & High                                                                          & V2G                                                         \\ \hline
Liwang's \cite{8489932}      & reverse                                                           & $\checkmark$                                                              & $\times$                                                                          & $\times$                                                                  & $\times$                                                                  & $\times$                                                             & $\checkmark$                                                          & Low                                                                      & High                                                                          & \begin{tabular}[c]{@{}c@{}}vehicular comp.\\ offloading\end{tabular} \\ \hline
BidGuard \cite{8169049}     & single & $\checkmark$                                                              & $\times$                                                                          & $\checkmark$                                                              & $\times$                                                                  & $\times$                                                             & $\times$                                                              & Low                                                                      & Low                                                                           & MCS                                                       \\ \hline
Wang's \cite{8737594}      & reverse                                                           & $\checkmark$                                                              & $\times$                                                                          & $\checkmark$                                                              & $\times$                                                                  & $\times$                                                             & $\times$                                                              & Low                                                                      & Low                                                                           & MCS                                                       \\ \hline
{Trustee \cite{Trustee2020Galal}}      & {Vickrey}                                                 & {$\checkmark$}                                                              & {$\times$ }                                                                         & {$\checkmark$}                                                              & {$\checkmark$ }                                                                 & {$\times$ }                                                            & {--- }                                                             & {Medium }                                                                     & {High }                                                                          & {General }                                                      \\ \hline
{SAFE \cite{9296813}}      & {single-round}                                                 & {$\checkmark$}                                                              & {$\times$}                                                                          & {$\checkmark$}                                                              & {$\checkmark$}                                                                  & {$\checkmark$}                                                             & {--- }                                                             & {Medium }                                                                     & {High }                                                                          & {General }                                                      \\ \hline
\textbf{SEAL} & \begin{tabular}[c]{@{}c@{}}SRC\end{tabular}   & $\checkmark$                                                              & $\checkmark$                                                                      & $\checkmark$                                                              & $\checkmark$                                                              & $\checkmark$                                                         & $\checkmark$                                                          & Low                                                                      & High                                                                          & \begin{tabular}[c]{@{}c@{}}UAV comp.\\ offloading\end{tabular}       \\ \hline
\multicolumn{11}{l}{Note 1: ``$\checkmark$" means support; ``$\times$" means not support; ``---" means not mentioned; ``comp." means computation; ``comm." means communication.}\\
\multicolumn{11}{l}{{Note 2: ``Comp.\&Comm. Overhead" is evaluated under our UAV computation offloading scenario with high-frequency resource trading.}}
\end{tabular} }\vspace{-0.25cm}
\end{table*}

\section{Related Works}\label{sec:RELATEDWORK}
\subsection{Computation Offloading Mechanisms for UAVs}\label{subsec:relatedwork1}
{Various efforts on computation offloading have been made in wireless networks.
Xiong \emph{et al}. \cite{8470083} model the offloading of proof-of-work (PoW) tasks from resource-limited miners to cloud/fog servers as a Stackelberg game and analyze the Stackelberg equilibrium.
Ng \emph{et al}. \cite{9629303} present a double auction to match vehicles' required computation resources with edge servers under coded distributed computing paradigms.
Gao \emph{et al}. \cite{9355000} propose a truthful auction mechanism to offload graph jobs efficiently under the vehicular cloud computing paradigm.
The preceding works, however, are inapplicable to UAV networks with 3D mobility and complex aerial-ground dynamics. Furthermore, neither fairness nor the preservation of entities' privacy are taken into account.}

Recently, a mass of works have been reported on computation offloading mechanisms for energy-limited UAVs.
By formulating the dynamic offloading problem as a Markov process, Callegaro \emph{et al}. \cite{8648099} derive the optimal policies for UAVs to partially offload the computation missions to urban fog nodes with consideration of node competition and server congestion.
Bai \emph{et al}. \cite{8693989} investigate a fog computing-assisted efficient task offloading mechanism for UAV swarms to extend their battery recharging time under partial and full offloading scenarios.
Hou \emph{et al}. \cite{8761932} study the optimal computation task assignment problem for UAV fleets under urban fog computing environment and design a genetic algorithm to obtain the near-optimal solution for energy consumption minimization.
Liwang \emph{et al}. \cite{9453820} design a novel futures trading paradigm for onsite resource trading between UAVs and ground edge servers to relieve trading failures, latency, and unfairness. They also present two algorithms for optimal forward contract design and transmit power optimization.
Sacco \emph{et al}. \cite{9693227} investigate a feasible reinforcement learning based method to offload UAVs' computation tasks to ground edge clouds. 

Distinguished from the above works on UAV computation offloading to fixed fog servers or remote cloud servers, our work aims to design an efficient and on-demand scheme by harnessing idle computing resource shared by ground vehicles (referred to as VFC) to offload heavy computation missions produced by UAVs.
{In our previous work \cite{9696188}, we study a VFC-based task offloading framework for UAVs in disaster scenarios and design a stable one-to-one matching algorithm for task scheduling. Nevertheless, the fairness and privacy issues in task scheduling are ignored in \cite{9696188}.}

\subsection{Strategy-Proof and Privacy-Preserving Auction Mechanisms}\label{subsec:relatedwork2}
There have been a number of recent efforts on strategy-proof and privacy-preserving auction mechanisms, which mainly depend on advanced cryptographic mechanisms (e.g., zero-knowledge proof (ZKP) and homomorphic encryption (HE)) and differential privacy (DP).
Chen \emph{et al}. \cite{8493354} propose ARMOR which leverages cryptographic tools including HE and garbled circuits to protect users' location and bid information in combinatorial spectrum auctions while considering spectrum reusability.
Wang \emph{et al}. \cite{8675981} present a secure double auction named PS-TAHES for spectrum redistribution based on HE and garbled circuits to enable privacy-preserving bid multiplication, comparison, and sorting matrix operations.
Wan \emph{et al}. \cite{9354536} develop V2GEx, a blockchain system with ZKP support to ensure exchange fairness and payment privacy for electric vehicles in vehicle-to-grid (V2G) energy services.
Nevertheless, these approaches based on advanced cryptography may introduce large system overheads and consume considerable resource for energy-limited UAVs.
Lin \emph{et al}. \cite{8169049} present a differentially private truthful auction architecture named BidGuard in mobile crowdsensing (MCS) to prevent adversaries from deducing users' private information via inference attacks.
Wang \emph{et al}. \cite{8737594} propose a differentially private reverse auction framework for MCS under an untrusted auctioneer, where the exponential mechanism is employed to locally obfuscate users' bids to prevent inference attacks.
However, these solutions built on DP may result in a large bid utility decrease for practical use during the perturbation process, thereby {deteriorating} the auction efficiency in winner and payment determination.

{Recently, there has been a surge in interest in combining smart contract and trusted execution environment (TEE) technologies to protect the bid privacy in auctions \cite{Trustee2020Galal,9049585,WANG2021223secure,Shi2021When,9296813}. Galal and Youssef \cite{Trustee2020Galal} develop Trustee to fully preserve bid privacy in Vickrey auctions by integrating the Ethereum and an Intel SGX enclave. 
In \cite{Trustee2020Galal}, bidders send their encrypted bids to the smart contract within the bidding interval, and the enclave executes the auction program and produces a signed transaction (including winners and payments) to the smart contract. Brandenburger \emph{et al}. \cite{9049585} identify that smart contracts run inside TEEs are vulnerable to rollback attacks and present a secure architecture implemented by Hyperledger Fabric for smart contract execution within Intel SGX with rollback prevention. Wang \emph{et al}. \cite{WANG2021223secure} leverage the smart contract to replace the untrusted auctioneer to run spectrum auctions and utilize Intel SGX and Pedersen commitment to preserve bid privacy for public verification in smart contracts. 
Chen \emph{et al}. \cite{9296813} develop a general fair and privacy-preserving auction framework named SAFE based on smart contracts and Intel SGX, where four representative single-round auction formats are utilized as examples to show the framework design.}

{However, the above works \cite{Trustee2020Galal,9049585,WANG2021223secure,Shi2021When} do not consider the threats arising from the exchange fairness in the auction, which may discourage honest bidders from truthfully participating in auctions. Although the work \cite{9296813} considers the exchange fairness in system design, it primarily applies to general single-round auctions and can result in high system costs (e.g., high Gas fee) for multi-round auctions, particularly for UAV computation offloading scenarios with highly frequent task offloading and highly dynamic network environment. 
In opposite, our SEAL implements a fair exchange protocol including a hashchain-based batch payment mechanism (to reduce operational cost of smart contracts) and a commit-then-claim mechanism (to coordinate on-chain fair exchange and off-chain payword delivery) for UAV computation offloading applications. Besides, for heterogeneous UAV computation tasks, we design a novel multi-round SRC auction with combinatorial truthfulness guarantee.} Our SEAL ensures a wide range of security targets in a trust-free manner with much {improved} efficiency than its alternatives.
A comparison of our work with other competing approaches is given in Table~\ref{tableCMP}.

\section{System Model}\label{sec:SYSTEMMODEL}
In this section, we describe the system model by discussing the network model, mobility model, auction model, threat model, and design goals. Table \ref{notationtable} summarizes the key notations in this paper.

\begin{table}\scriptsize
\setlength{\abovecaptionskip}{-0.08cm}
\caption{Summary of Notations}\label{notationtable}\centering
\begin{tabular}{|c|l|}
\hline
\textbf{Notation} & \textbf{Description} \\  \hline 
    $\mathbb{I}$&Set of ground vehicles serving as VFC nodes. \\
    $\mathbb{J}_n$&Set of computation tasks at location $n$. \\
    $\mathbb{W}$&Set of winners of task allocation. \\
    $\mathbb{G}_{i}$&Set of allocated tasks to winner $i$. \\
    $\Im_{j,n}$&UAV's $j$th task at location $n$. \\
    $\Re$&Radius of A2G/G2A communication coverage of UAV. \\
    $s_{j,n}$&Data size of task $\Im_{j,n}$. \\
    $\varphi_{j,n}$&Task urgency degree or processing priority. \\
    $\tau_{j,n}$&Task completion deadline.    \\
    $\zeta_{j,n}$&Computing intensity. \\
    $\hbar_n$&Flying altitude of UAV at location $n$. \\
    $K$&Number of evenly divided time slots with interval $\Delta_t$. \\
    $V_n[t_k]$&UAV's flying speed at time slot $k$ at segment $n$. \\
    $\overline \vartheta_{\mathrm{veh}}$&Average vehicular speed.  \\
    $\mathcal{B}_i$&Combinatorial bid of vehicle $i$. \\
    $\Gamma_{i}$&Feasible task bundle of vehicle $i$. \\
    $\chi_{i}^{j}$&Amount of computing resources of vehicle $i$ in task $\Im_{j,n}$. \\[0.2pt]
    ${b_{i}^{j}}$&Bidding price of vehicle $i$ for task $\Im_{j,n}$. \\[0.2pt]
    $\beta_{i}^{j}$&Binary task allocation variable. \\
    $p_{i}^{j}$&Payment to winner $i$ for $j$th task. \\
    $\Theta({\chi_{i}^{j}})$&Cost of vehicle $i$ in task offloading. \\
    $E_n$&Energy consumption of the UAV at location $n$. \\
    $\varpi$&Weight factor of UAV's energy cost and payment. \\
    ${T_{j,n}}$&Task completion time of mission $\Im_{j,n}$. \\
    $\tau_i^R$&Residual dwell time of vehicle $i$ in UAV's coverage. \\
    $\mathbb{C}_{j,n}$&Feasible candidate set of task $\Im_{j,n}$. \\
    $\digamma(\chi _{i}^{j},b_{i}^{j})$&Marginal cost factor of vehicle $i$. \\[2.pt]
    $\widetilde{b}_{i,j}$&Virtual bidding price of vehicle $i$. \\
\hline
\end{tabular}\vspace{-0.2cm}
\end{table}

\subsection{Network Model}\label{subsec:networkmodel}
Fig.~\ref{fig:systemmodel} illustrates a typical scenario where a UAV is dispatched for data sensing missions along a pre-determined transit route containing $N$ locations of interest.
The UAV starts from location $1$, then sequentially moves to the next location by following the straight-line trajectory \cite{9043589}, and finally ends at location $N$.
When the UAV arrives at the sensing location $n$, it {hovers over this location to capture the sensory data (e.g., videos and images) of the task area with onboard sensors}, then it generates a set of $\mathbb{J}_n=\{1,\cdots,j,\cdots,J_n\}$ computing missions (e.g., image and video processing).
For improved endurance capability, these missions can be offloaded to VFC nodes (i.e., ground vehicles with idle computing resources) for processing via air-to-ground (A2G) links, and the processing results are sent back to the UAV via ground-to-air (G2A) links\footnotemark[1]. \footnotetext[1]{When vehicles are not available in the UAV's communication range or the computing resources of vehicles are not sufficient, UAV's computation tasks can be alternatively offloaded to the remote cloud, as in conventional works \cite{8981670,8723343}.}The radius of A2G/G2A communication coverage of the UAV is denoted as $\Re$.
Here, each mission $j \in \mathbb{J}_n$ can be described by a $4$-tuple:
\begin{align}\label{eq:task}
\Im_{j,n} = \left\langle s_{j,n}, \varphi_{j,n}, \tau_{j,n}, \zeta_{j,n} \right\rangle, 1\le j \le J_n,
\end{align}
where $s_{j,n}$ (in bits) is the task size, $\varphi_{j,n}$ is the urgency degree indicating the priority for task processing, $\tau_{j,n}$ (in seconds) is the task completion deadline, and $\zeta_{j,n}$ (in CPU cycles/bit) is the computing intensity.

\subsection{Mobility Model}\label{subsec:mobilitymodel}
During the execution of task $\Im_{j,n}$, the UAV is assumed to hover in the sky at the constant flying altitude $\hbar_n$ to avoid frequent ascending and descending for minimized energy consumption, where $\hbar_n$ is the minimum altitude suitable for the working area and can avoid all collisions and blockages \cite{9043589}. {Thereby, we only have to consider the mobility of vehicles in each sensing location.} 
For simplicity, the time horizon $T$ is evenly divided into $K$ time slots \cite{9696188}, and each time slot has an interval of $\Delta_t = \frac{T}{K}$. Let $V_n[t_k]$ be the UAV's flying speed at segment $n$ between any two successive locations $n$ and $n+1$ ($n,n+1 \leq {N}$) at time slot $k$.
The set of ground vehicles serving as VFC nodes in the UAV's coverage at location $n$ is denoted as $\mathbb{I}=\{1,\cdots,i,\cdots,I\}$.
Specifically, {the vehicle flow (i.e., average number of vehicles entering the UAV's coverage unit time) is denoted as $\lambda$, which is assumed to follow a Poisson process with mean arrival rate $\lambda$ \cite{5678795}. According to \cite{5678795,fricker2004fundamentals}, we have $\lambda = \eta\, \overline \vartheta_{\mathrm{veh}}$,} where $\eta$ (in veh/km) is the vehicle density\footnotemark[2]\footnotetext[2]{{In this paper, for ease of analysis, vehicles are driving on the straight road beneath the UAV. The vehicle density in the UAV's coverage is computed as $\eta = I/L$, where $I$ means the number of vehicles on the road segment in the UAV's coverage and $L$ is the length of the road segment in the UAV's coverage.}} and $\overline \vartheta_{\mathrm{veh}}$ is the average vehicle speed.
{According to field observations in \cite{fricker2004fundamentals}, the vehicular speed-traffic relationship can be described as\cite{5678795,6180101}:}
\begin{align}\label{eq:velocityGV}
\overline \vartheta_{\mathrm{veh}} = \max \left\{ {{\vartheta_{\mathrm{veh}}^{\min }},\left( {1 - {{\eta}}/{{\eta _{\max }}}} \right){\vartheta_{\mathrm{veh}}^{\max }}} \right\},
\end{align}
where ${\eta _{\max }}$ means the maximum traffic density {(i.e., vehicle jam density at which traffic comes to a halt \cite{5678795,fricker2004fundamentals,6180101})}. ${\vartheta_{\mathrm{veh}}^{\min }}$ and ${\vartheta_{\mathrm{veh}}^{\max }}$ are the minimum and maximum vehicle speed, respectively. {As seen in Eq. (\ref{eq:velocityGV}), when vehicle density $\eta$ grows from zero, the vehicle flow $\lambda$ on the road also grows while the vehicle speed $\overline \vartheta_{\mathrm{veh}}$ declines (referred to as the free-flow phase). When the density reaches or above its threshold (i.e., $\eta \ge \left( {1 - {\vartheta_{\mathrm{veh}}^{\min }}/{\vartheta_{\mathrm{veh}}^{\max }}} \right){\eta _{\max }}$), the vehicle traffic becomes congested, which entails low vehicle speed (referred to as the congested-flow phase).}

Here, the number of vehicles entering the UAV's coverage at time slot $k$ is
$\mu^{\mathrm{in}}[t_k] = \lambda \Delta_t = \eta\, \overline \vartheta_{\mathrm{veh}} \Delta_t.$
Let $\mu^{\mathrm{out}}[t_k]$ denote the ratio of vehicles leaving the UAV's coverage. Then, the number of vehicles in the UAV's communication range at time slot $k$ is denoted as:
\begin{align}\label{eq:vehiclenumber}
{I}[t_k] = \begin{cases}
\left(\mu^{\mathrm{in}}[t_k] + {I}[t_{k-1}]\right) \left(1 - \mu^{\mathrm{out}}[t_k]\right),&1 < k \le K;\\
~\mu^{\mathrm{in}}[t_1] (1 - \mu^{\mathrm{out}}[t_1]), &k=1.
\end{cases}
\end{align}

\subsection{Auction Model}\label{subsec:auctionmodel}
The auction model is employed to schedule UAV's computation task offloading, where ground vehicles are the bidders while the UAV severs as the auctioneer.
Each bidder $i\in \mathbb{I}$ submits its combinatorial bid $\mathcal{B}_i=\Big\langle \Gamma_{i}, \overrightarrow{\chi_{i}}, \overrightarrow{b_{i}} \Big\rangle$ including the feasible task bundle $\Gamma_{i}$, the computing resource profile $\overrightarrow{\chi_{i}}$, and the bidding price profile $\overrightarrow{b_{i}}$. Here, $\overrightarrow{\chi_{i}}=\{{\kappa_{i}^{j}\chi_{i}^{j}}\}_{j=1}^{\left|\Gamma_{i} \right|}$, $\overrightarrow{b_{i}}=\{{b_{i}^{j}}\}_{j=1}^{\left|\Gamma_{i} \right|}$, $\kappa_{i}^{j}\in(0,1]$. $b_{i}^{j}\ge 0$ is vehicle $i$'s bidding price\footnotemark[3] for task $\Im_{j,n}$.\footnotetext[3]{{The bidding price indicates the bidder's intended payment to be received from the UAV for completing the offloaded computation task. The bidding price is generally determined by the valuation of vehicle user (which is measured by the cost in contributing computation resources).}} For ${t_{i}^{j}} \notin \Gamma_{i}$, we have ${\chi_{i}^{j}}= 0$ and $b_{i}^{j}= \infty$. 
We use $\mathbf{b}_i^{j}$ to denote the combinatorial bid for $j$th task, i.e.,
\begin{align}\label{eq:bidderutility}
\mathbf{b}_i^{j}=({t_{i}^{j}}, {\kappa_{i}^{j}\chi_{i}^{j}}, {b_{i}^{j}}), \forall {t_{i}^{j}} \in \Gamma_{i}.
\end{align}
Then, $\mathcal{B}_i$ can be rewritten as $\mathcal{B}_i=\{ \mathbf{b}_i^{1}, \cdots,\mathbf{b}_i^{j}, \cdots,\mathbf{b}_i^{{\left|\Gamma_{i} \right|}} \}$.
In the auction, each vehicle $i$ bids ${b _{i}^{j}}$ to sell ${\kappa_{i}^{j}\chi_{i}^{j}}$ amount of computing resources for each task ${t_{i}^{j}} \in \Gamma_{i}$.
In addition, vehicles are supposed to be \emph{single-minded} \cite{NISAN2015477}, {indicating that} they can only sell the reported amount of computing resources (i.e., $\kappa_{i}^{j}=1$) or lose the auction.
\\
\textbf{Definition 1 (Single-minded Reverse Combinatorial (SRC) Auction).} \emph{Given the task set $\mathbb{J}_n$ and the bid profile $\mathcal{B} = \{\mathcal{B}_1,\cdots,\mathcal{B}_{I}\}$ with $\kappa_{i}^{j}=1$, a SRC auction $\mathcal{A}$ can be denoted as a pair of allocation rules $\overrightarrow{\beta}(\mathcal{B})$ and payment rules $\overrightarrow{p}(\mathcal{B})$.
Here, $\overrightarrow{\beta}(\mathcal{B})\!=\!(\beta_{i}^{j})_{\left|\mathbb{J}_n\right|\times \left|\mathbb{I}\right|}$, $\overrightarrow{p}(\mathcal{B}) \!=\!(p_{i}^{j})_{\left|\mathbb{J}_n\right|\times \left|\mathbb{I}\right|}$.
$\beta_{i}^{j}$ is a binary variable, where $\beta_{i}^{j}\!=\!1$ means bidder $i$ wins to execute $j$th mission and $\beta_{i}^{j}\!=\!0$ means bidder $i$ loses. $p_{i}^{j}$ is the payment to winner $i$ for $j$th task.}
\\
\textbf{Definition 2 (Payoff of ground vehicle).} \emph{The payoff {(or utility)} of each bidder $i\in \mathbb{I}$ (i.e., ground vehicle) for task $\Im_{j,n}$ is {the payment minuses its monetary resource cost \cite{8660495,7559774,7118253}, i.e.,}
\begin{align}\label{eq:bidderutility}
\pi_{i}^{j}=\pi({\chi_{i}^{j}},b_{i}^{j}) = \beta_{i}^{j}\left(p_{i}^{j}- \Theta({\chi_{i}^{j}})\right),
\end{align}
where $\Theta({\chi_{i}^{j}})$ is the cost of bidder $i$ in sharing ${\chi_{i}^{j}}$ amount of computing resource, which is private and unknown to others.}

\subsection{Threat Model}\label{subsec:threatmodel}
\begin{figure}[!t]\setlength{\abovecaptionskip}{-0.05cm}
\centering
  \includegraphics[width=8.2cm]{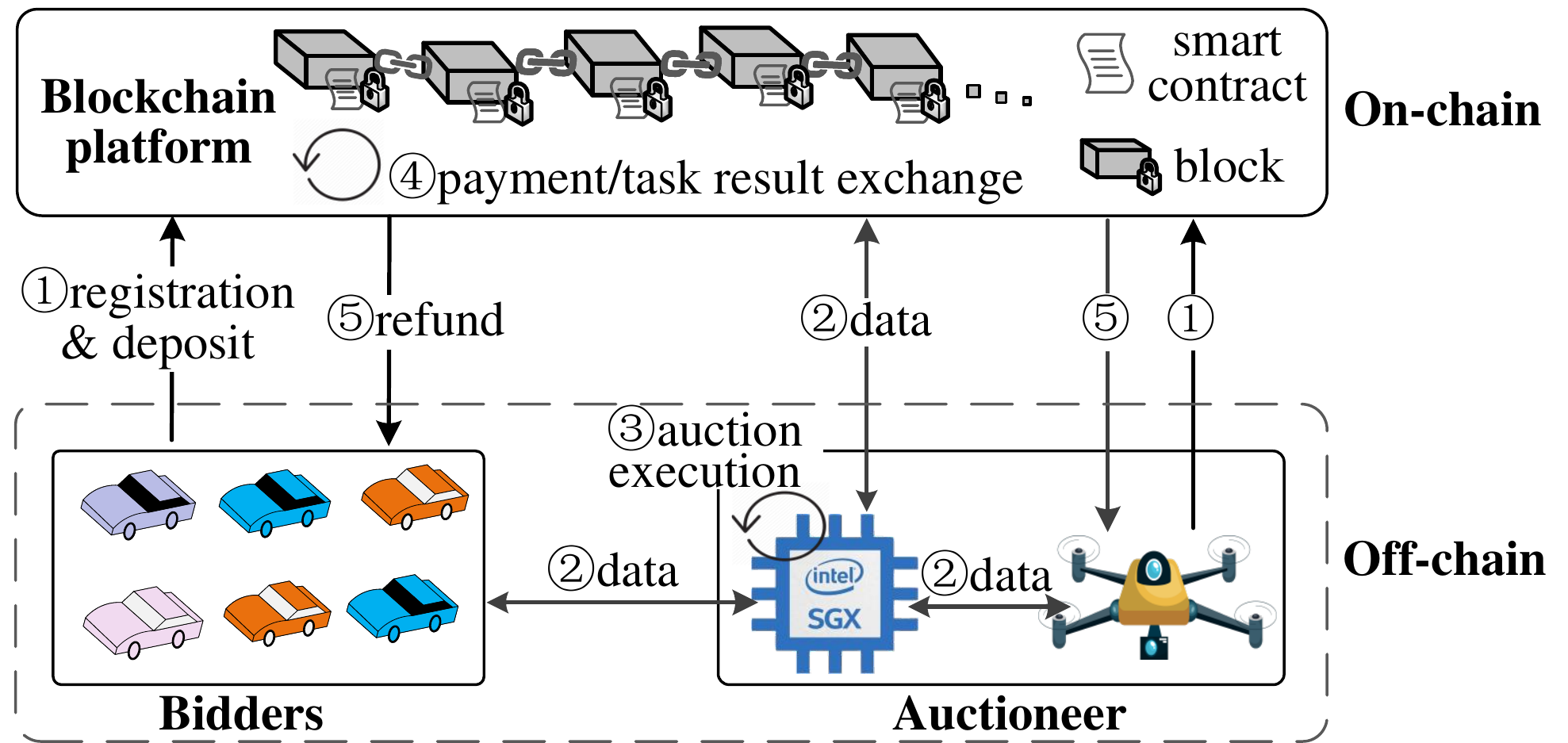}\\
  \caption{An illustration of SEAL system.} \label{fig:overview}\vspace{-0.3cm}
\end{figure}

In truthful auctions, the truthful bid information can divulge bidders' true valuations and resource costs. If the private bids are unauthorizedly exposed to the public, malicious bidders may take advantage of these information to seek higher profits and even {conduct} market manipulation in current or future auctions. For privacy preservation, the UAV is equipped with a trusted processor and it executes the bid collection and SRC auction process inside the TEE, as illustrated in Fig.~\ref{fig:overview}. The permissioned blockchain and smart contracts are exploited to improve the fairness of the exchange of payment and task result in the auction. The following security assumptions are made. 
\begin{itemize}
  \item \textbf{Bidders}: The bidders (i.e., ground vehicles) are assumed to be \emph{strategic and selfish} during task offloading. They may deviate from the auction protocol to increase their payoffs and even manipulate the auction by submitting untruthful bids and aborting the assigned tasks deliberately.
  \item \textbf{Auctioneer}: The auctioneer (i.e., the UAV) is assumed to be \emph{semi-honest} (i.e., \emph{passive}). More specifically, the UAV will honestly follow the auction protocol but is curious about participants' private bid information and may refuse to pay after receiving the computing results from vehicles.
  \item \textbf{TEE}: The TEE enclave (deployed on the auctioneer) implemented by Intel SGX is supposed to be \emph{secure}, and the correctness of its sealed data (e.g., bids and programs) can be verified via remote attestations \cite{Anati_SGX}. 
  \item \textbf{Blockchain}: A permissioned blockchain is maintained by all participants (i.e., UAVs and vehicles), and a secure consensus algorithm is assumed to be executed by participants for blockchain maintenance. The transactions recorded in hash-chained blocks are supposed to be tamper-resistant. 
\end{itemize}

\vspace{-2mm}
\subsection{Design Goals}\label{subsec:goals}
The goal of our SEAL is to achieve the following desirable properties simultaneously.

\textbf{1) Combinatorial strategy-proofness guarantee.} Strategy-proofness (or truthfulness) is the fundamental basis for an auction mechanism \cite{7559774}, whose formal definition is given as below.
\\
\textbf{Definition 3 (Combinatorial Strategy Proofness).} \emph{An auction mechanism is combinatorial strategy-proof if both combinatorial incentive compatibility and individual rationality are guaranteed.}
\begin{itemize}
  \item  Combinatorial incentive compatibility (CIC). \emph{An auction is  combinatorial incentive-compatible if reporting truthful combinatorial bid information ${\overrightarrow{\mathcal{B}}_{i}^{j}}^*=({\chi_{i}^{j}},\Theta({\chi_{i}^{j}}))$ is the dominant strategy for every bidder $i\in \mathbb{I}$ to maximize its payoff regardless of other bidders' strategy profile $\overrightarrow{\mathcal{B}}_{-i,j,n}$, i.e., $\pi({\overrightarrow{\mathcal{B}}_{i}^{j}}^*,\overrightarrow{\mathcal{B}}_{-i}^j)\geq \pi(\overrightarrow{\mathcal{B}}',\overrightarrow{\mathcal{B}}_{-i}^j)$, $\forall \overrightarrow{\mathcal{B}}' \ne {\overrightarrow{\mathcal{B}}_{i}^{j}}^*$.}
  \item Individual rationality (IR). \emph{An auction is individual-rational if the expected payoff of each bidder $i\in \mathbb{I}$ participating in the auction is no less than that under non-participation, i.e., $\pi_{i}^{j}\ge 0$, $\forall {j\in\Gamma_{i}}$.}
\end{itemize}

\textbf{2) Privacy preservation for bidders.} Different from most of the existing auction approaches \cite{7118253,8493354,8675981,9354536,8489932,8169049} that exploit heavy cryptographic tools (such as garbled circuits and HE), or integrate DP methods to preserve the bidding privacy, SEAL aims for an efficient privacy-preserving auction scheme in terms of system overheads and auction efficiency, by leveraging smart contracts with the aid of the trusted processor. 

\textbf{3) Fair exchange between distrustful participants.} Fairness is another essential target for auction mechanism design to prevent bidders' malicious dropout and eliminate the exchange dilemma. The definition of fairness in the auction is given as follows \cite{9296813}.
\\
\textbf{Definition 4 (Fairness).} \emph{A fair auction mechanism satisfies both participation fairness and exchange fairness.}
\begin{itemize}
  \item Participation fairness. \emph{An auction ensures participation fairness if any rational and selfish UAV is stimulated to honestly obey the auction protocols, i.e., they have no incentives to bid untruthfully or abort the auction.}
  \item Exchange fairness. \emph{An auction ensures exchange fairness if the exchange can be faithfully realized between mutually untrusted entities in the auction.}
\end{itemize}

\textbf{4) UAV cost minimization.} SEAL aims to minimize UAV's energy cost and payment in dynamic computation offloading environments with low computation and communication overheads.

%


\section{SEAL: Our Detailed Construction}\label{sec:FRAMEWORK}
\subsection{Framework Overview}\label{subsec:scheme0}
For UAV computation offloading services based on truthful auctions, the implementation of smart contracts requires the replication of all involved data (including the bid information that can reveal bidders' true types and valuations) to all participants for public audit and mutual supervision \cite{9268472,9928220}, thereby leaking bidders' privacy. Besides, the high frequency of computation offloading behaviors and the corresponding huge number of micropayments may deteriorate the performance of the smart contract system. We make the following two improvements in SEAL to address these two challenges. 
One is to move the computation of auction process into an off-chain trusted processer by introducing the concept of local consensus; the other is the hashchain-based batch payment protocol to reduce the cost {of} supporting frequent on-chain payments. At a high level, as shown in Fig.~\ref{fig:overview}, the design of SEAL orchestrates two parts: \emph{on-chain} and \emph{off-chain}.
\begin{itemize}
  \item The \textbf{on-chain} part automatically executes the auction-based offloading services (i.e., automated delivery of computing results and {due} payment) by the smart contract with fairness and transparency guarantees, where the service transactions are immutably stored on the blockchain ledgers for audit.
  \item The \textbf{off-chain} part implements the strategy-proof SRC auction mechanism into the TEE to guarantee {the} auction correctness without violating bidders' privacy. Besides, the UAV makes micropayment commitments based on {the} hashchain and sequentially sends them to bidder vehicles as the payment authorization after seeing the proof-of-publication of task result delivery on the blockchain, and then the bidder claims the due payment based on the received commitments.
\end{itemize}

\begin{figure}[!t]
\centering\setlength{\abovecaptionskip}{-0.05cm}
  \includegraphics[width=9.3cm]{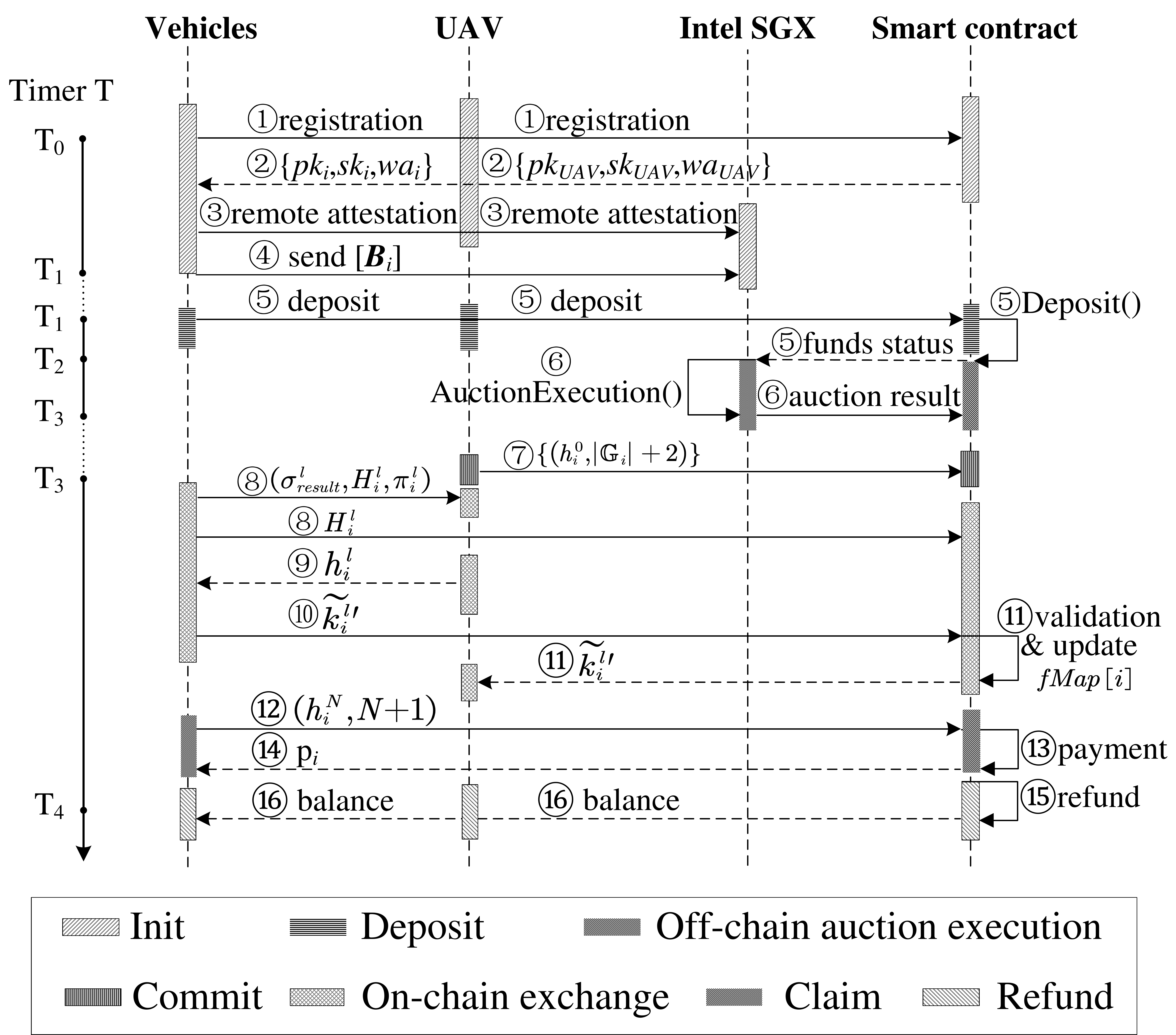}\\
  \caption{Sequence diagram of SEAL.} \label{fig:workflow}\vspace{-0.26cm}
\end{figure}

Concretely, as illustrated in Fig.~\ref{fig:workflow}, there are 8 phases need to undertake in the design of SEAL, i.e., \emph{Init}, \emph{Deposit}, \emph{Off-chain auction execution}, \emph{Commit}, \emph{On-chain exchange}, \emph{Claim}, \emph{Refund}, and an additional \emph{Timeout} operation.
The \textbf{init} phase (step \ding{172}--\ding{175}) executes on-chain registration and off-chain initialization. The TEE performs \textbf{off-chain auction execution} operations (step \ding{177}) using the local consensus to privately produce auction results after validating user deposits in the \textbf{deposit} phase (step \ding{176}). The fair exchange is realized by the commit-then-claim mechanism and \textbf{on-chain exchange} operations (step \ding{179}--\raisebox{.9pt}{\small{\textcircled{\scriptsize{\textbf{11}}}}}), where the auctioneer makes hashchain-based micropayments in the \textbf{commit} phase (step \ding{178}) and sequently sends micropayments to the bidder as the payment authorization after on-chain task result delivery, and the bidder can claim the due payment in the \textbf{claim} phase (step \raisebox{.9pt}{\small{\textcircled{\scriptsize{\textbf{12}}}}}--\raisebox{.9pt}{\small{\textcircled{\scriptsize{\textbf{14}}}}}). After that, the participant can obtain the remaining fund in the \textbf{refund} phase (step \raisebox{.9pt}{\small{\textcircled{\scriptsize{\textbf{15}}}}}--\raisebox{.9pt}{\small{\textcircled{\scriptsize{\textbf{16}}}}}).

\subsection{Smart Contract Execution with TEE}\label{subsec:scheme1}

\begin{algorithm}[t!]\begin{small}
   \caption{{Smart Contract with Off-Chain Auction Execution}}\label{Algorithm1}
    \begin{algorithmic}[1]
        \STATE \textbf{\underline{Init}:}
        \STATE Check the timer $T$. If $T_0 < T < T_1$, proceed to line 3, otherwise go to line 5;
        \STATE $\mathrm{bidder}~i\leftarrow\{pk_i,sk_i,wa_i\}$, $\mathrm{UAV}\leftarrow\{pk_{\mathrm{UAV}},sk_{\mathrm{UAV}},wa_{\mathrm{UAV}}\}$;
        \STATE Each bidder $i$ sends $[\mathcal{B}_i]\leftarrow\mathbf{Enc}_{pk_{\mathrm{TEE}}}(\mathcal{B}_i)$ to the TEE.
        \STATE If $T_1 < T < T_2$, proceed to line 6, otherwise terminate;
        \STATE Each seller and buyer invokes \textbf{Deposit()}.
        \STATE \textbf{\underline{Deposit}:}
        \STATE Each participant $m$ sends $(\mathrm{deposit}, \$val_m)$ to the smart contract.
        \IF{$balance_m\ge \$val_m$}
        \STATE Smart contracts transfer $\$val_m\rightarrow esPool$ from $wa_m$;
        \STATE Smart contracts update $Depo[m]\leftarrow \$val_m$;
        \ENDIF
        \STATE \textbf{\underline{Off-Chain Auction Execution}:} \textcolor{gray}{\#\,Called by the TEE.\,\#}
        \STATE If $T_2 < T < T_3$, proceed to line 15, otherwise terminate;
        \STATE Decrypt $\mathcal{B}_i\leftarrow\mathbf{Dec}_{sk_{\mathrm{TEE}}}([\mathcal{B}_i])$;
        \STATE Check the deposit $Depo[m]$ of each bidder and the UAV;
        \STATE Decide the winners and payments and publish $(\overrightarrow{\beta},\overrightarrow{p})$ on blockchain.
        \STATE If $T_3 < T < T_4$, proceed to line 19, otherwise terminate;
        \STATE \textbf{\underline{Commit}:}
        \STATE The UAV generates a hashchain $HashChain_i$ of length $\left|\mathbb{G}_i \right|+2$ for each winner $i\in \mathbb{W}$;
        \STATE The UAV sends the signed metadata $\{(h_i^0,\left|\mathbb{G}_i \right|+2)\}_{i\in \mathbb{W}}$ to the smart contract.
        \STATE \textbf{\underline{On-Chain Exchange}:}
        \STATE \textcolor{gray}{\#\,Each winning bidder $i$ executes lines 24-25 for $l\!=\!1,2,\cdots,\left|\mathbb{G}_i\right|$.\,\#}
        \STATE Compute $\sigma_{result}^l\leftarrow\mathbf{Enc}_{pk_{\mathrm{UAV}}}(\mathbf{Enc}_{k_i^l}(result_i^l))$, $H_i^l\leftarrow\mathbf{Hash}(k_i^l||nonce_l)$, and a zero knowledge proof $\pi_i^l$;
        \STATE Send $(\sigma_{result}^l, H_i^l,\pi_i^l)$ to the UAV and publish $H_i^l$ on the blockchain.
        \STATE \textcolor{gray}{\#\,The UAV executes lines 27-31.\,\#}
        \IF{$\mathbf{Verify}(\pi_i^l)=true$}
        \STATE Sequentially transmit the signed hash value $h_i^l$ of $HashChain_i$;
        \ELSE
        \STATE Report the misbehavior of the bidder $i$ to the smart contract;
        \ENDIF
        \STATE Winner $i$ sends $\widetilde{k_i^l}'\leftarrow \mathbf{Enc}_{sk_i}({k_i^l}'||nonce_l)$ to the smart contract. 
        \STATE \textcolor{gray}{\#\,The smart contracts execute lines 34-39.\,\#}
        \STATE Decrypt $({k_i^l}',nonce_l)\leftarrow \mathbf{Dec}_{pk_i}(\widetilde{k_i^l}')$;
        \IF{$\mathbf{Hash}({k_i^l}'||nonce_l) = H_i^l$ \& $T\leq \tau_{l,n}$}
        \STATE Deliver the key ${k_i^l}'$ to the UAV;
        \ELSE
        \STATE Update the failed task map of winner $i$ as $fMap[i]\leftarrow\{l\}$;
        \ENDIF
        \STATE The UAV obtains $result_i^l \leftarrow\mathbf{Dec}_{{k_i^l}'}(\mathbf{Dec}_{sk_{\mathrm{UAV}}}(\sigma_{result}^l))$.
        \STATE \textbf{\underline{Claim}:}
        \STATE Winner $i$ sends $(h_i^N,N\!+\!1)$ to the smart contract for validation;
        \STATE Smart contracts compute $p_{i}=\sum_{l=1}^{N}{p_{i}^l} - \sum_{k \in fMap[i]}{p_{i}^k}$;
        \STATE Smart contracts transfer $p_{i}\rightarrow wa_i$ from $esPool$;
        \STATE Smart contracts update $Depo[\mathrm{UAV}] \leftarrow Depo[\mathrm{UAV}]-p_{i}$.
        \STATE \textbf{\underline{Refund}:}
        \STATE Participant $m$ sends $(\mathrm{refund}, \$val_m')$ to the smart contract.
        \IF{$Depo[m] = \$val_m'$}
        \STATE Smart contracts transfer $\$val_m'\rightarrow wa_m$ from $esPool$;
        \STATE Smart contracts update $Depo[m] \leftarrow Depo[m]-\$val_m'$;
        \ENDIF
    \end{algorithmic}\end{small}
\end{algorithm}

Algorithm~\ref{Algorithm1} summarizes the workflow of smart contract execution with the following eight phases.

\textbf{Init} (lines 2--6). In the on-chain part, after membership registration at the certificate authority (CA), each registered entity $m$ (i.e., UAV and vehicle) maintains an account $account_m$ including its public/private key-pair $(pk_m,sk_m)$, wallet address $wa_m$, and certificate $Cer_m = \mathbf{Enc}_{sk_{\mathrm{CA}}}(pk_m||T_{stamp}||T_{\mathrm{exp}})$ in the permissioned blockchain network. Here, $\mathbf{Enc}(\cdot)$ is the encryption function, ${sk_{\mathrm{CA}}}$ is the secret key of the CA, $T_{stamp}$ is the timestamp of certificate creation, and $T_{\mathrm{exp}}$ is the expiration time.
In the off-chain initialization, each bidder can serve as the challenger and verify the correctness of the loaded auction program in the TEE via remote attestations.
After successful remote attestations, each bidder $i$ submits its sealed bid information $[\mathcal{B}_i]=\mathbf{Enc}_{pk_{\mathrm{TEE}}}(\mathcal{B}_i)$ to the TEE before the auction. $pk_{\mathrm{TEE}}$ is the public key of TEE.

\textbf{Deposit} (lines 8--12). Each bidder $i$ sends a deposit transaction to the smart contract:
\begin{align}\label{eq:depositTx}
\mathrm{tx_{deposit}} = \left<\mathrm{deposit},\$val_m, Depo[m],pk_m,T_{\mathrm{stamp}},\sigma_{\mathrm{d}}\right>,
\end{align}
where $\$val_{i}$ is the deposit value sent to the blockchain, $Depo[m]$ is node $m$'s deposit record, $T_{\mathrm{stamp}}$ is the timestamp for transaction creation, and $\sigma_{\mathrm{d}}$ is the signature on the hash digest of $\mathrm{tx_{deposit}}$.
After checking the account balance $balance_i$, the deposit $\$val_{i}$ is transferred from $account_i$ to the escrow pool $esPool$.

\textbf{Off-chain auction execution} (lines 14--18). In this phase, a critical challenge is the correctness of smart contract execution within TEE. Traditional smart contracts that require the acknowledgment of all entities may contradict the privacy targets if users' private bids are publicly accessible. We observe that the correctness of off-chain auction execution only matters to the contracting parties. Instead, we introduce the \emph{local consensus} by narrowing the universal consensus only to contracting parties, where each involved party in the auction can authenticate the loaded program and data via software attestations.

Concretely, the TEE first decrypts the received encrypted combinatorial bids from all bidders by using its secret key ${sk_{\mathrm{TEE}}}$, i.e., $\mathcal{B}_i=\mathbf{Dec}_{sk_{\mathrm{TEE}}}([\mathcal{B}_i])$.
Then, each participant in the auction can verify the correctness and authenticity of the loaded program and bid information in the TEE enclave by software attestations via the EPID protocol.
Only if all involved parties reach an agreement on the loaded auction program and data within the TEE enclave, the off-chain auction execution will continue.
During off-chain auction execution, the TEE verifies the deposit value of each participant, and then decides winners $\overrightarrow{\beta}$ and payments $\overrightarrow{p}$ based on the private input (i.e., users' bid information) and the strategy-proof SRC auction program (i.e., Algorithm~\ref{Algorithm2} in Sect.~\ref{subsec:scheme3}).
Next, the auction results $(\overrightarrow{\beta},\overrightarrow{p})$ are uploaded and immutably stored in the blockchain ledgers. Based on $\overrightarrow{\beta}$, both the winner set $\mathbb{W}$ and the set of allocated tasks $\mathbb{G}_i$ to each winner $i*\in \mathbb{W}$ can be computed.

Another critical issue is to ensure the atomicity of {transactions}, i.e., either both the payment (to vehicles) and the release of task results (to the UAV) are completed simultaneously, or none of them at the cost of their deposits. By using the \emph{commit-then-claim} mechanism and \emph{hashchain micropayment} method, we develop a fair exchange protocol with atomic completion guarantees and batch payment functions in smart contracts, containing the commit phase, on-chain exchange phase, and claim phase, as shown in lines 20--45 in Algorithm~\ref{Algorithm1}.

\textbf{Commit} (lines 20--21). For every winner $i\in \mathbb{W}$, the UAV produces a hashchain with length $\left|\mathbb{G}_i \right|+2$ and root $h_i^0$, i.e.,
\begin{align}\label{eq:hashchain}
HashChain_i = \{h_i^{\left|\mathbb{G}_i \right|+1}\rightarrow h_i^{\left|\mathbb{G}_i \right|}\rightarrow \cdots h_i^1\rightarrow h_i^0\},
\end{align}
where ${\left|\mathbb{G}_i \right|}$ is the total number of allocated tasks to winner~$i$.
In conventional hashchain approaches \cite{Rivest1997PayWord,8998372}, the root $h_i^0$ is public and any element $h_i^z, \forall z\geq 1$ satisfies $h_i^z = \mathbf{Hash}(h_i^{z+1})$. 
Thereby, any element $h_i^z, \forall z\geq 1$ can be employed as the commitment for a constant micropayment unit, and the due payment can be validated by revealing the received commitments and calculated based on the number of received commitments, which is more efficient than those using signatures for verification.

However, in conventional hashchain-based micropayment methods \cite{Rivest1997PayWord,8998372}, each element in the hashchain represents a commitment for a fixed task payment, which is only applicable for services with constant micropayment.
In our work, due to the heterogeneity of tasks in terms of required computation resources and task deadline, the rewards (i.e., payments) for executing distinct tasks are different. Here, we design a novel hashchain micropayment method under the heterogeneous task setting.
Specifically, for the first element, we have $h_i^0=\mathbf{Hash}(h_i^{1})$. For the remaining elements with $z\geq 1$, we further consider the heterogeneous payment in designing the hashchain, i.e., $h_i^z = \mathbf{Hash}(h_i^{z+1}||p_{i}^z)$. $p_{i}^z$ is the micropayment to winner $i$ for task $\Im_{z,n}$, $\forall z \in \mathbb{G}_i$.
The element in the hashchain is summarized as:
\begin{align}\label{eq:heading}
h_i^z=\left\{ \begin{array}{ll}
	\mathbf{Hash}(h_i^{z+1}||p_{i}^z),&1\le z\le \left|\mathbb{G}_i \right|;\\[2pt]
	\mathbf{Hash}(h_i^{1}),&z=0.\\
\end{array} \right.
\end{align}
After the construction of all the hashchains for winners (i.e., $HashChain_i, i\in \mathbb{W}$), the UAV publishes the following claim transaction and delivers it to the smart contract:
\begin{align}\label{eq:commitTx}
&\mathrm{tx_{commit}} = \\
&\resizebox{1.0\hsize}{!}{$\left<\mathrm{commit},(meta_i, \{p_{i}^z\}_{z=1}^{\left|\mathbb{G}_i \right|},pk_i)_{i\in \mathbb{W}}, pk_{\mathrm{UAV}}, T_{\mathrm{stamp}}, T_{\mathrm{exp}}, \sigma_{\mathrm{com}}\right>,$} \nonumber
\end{align}
where $meta_i=\{h_i^0,\left|\mathbb{G}_i \right|+2\}$ is the metadata of the hashchain, $T_{\mathrm{exp}}$ is the expiry time, and $\sigma_{\mathrm{com}}$ is the signature on the hash digest of $\mathrm{tx_{commit}}$.
Once vehicle $i$ completes the computation task $\Im_{z,n}$ ($1\le z\le \left|\mathbb{G}_i \right|$), the UAV sequentially transmits the (signed) hash value $h_i^{z+1}$ on the hashchain to bidder $i$, as a verifiable commitment for the micropayment $p_{i}^z$ in task $\Im_{z,n}$. 
Then, every winner $i$ can use the received hash values in its hashchain as payment authorizations to claim its due {reward} in conducting the tasks in $\mathbb{G}_i$ (as analyzed in the Claim phase).

\emph{\textbf{Remark}.} As this exchange process is performed off-chain, there are no intermediate transactions to be processed by the blockchain. Besides, only the final payment between the UAV and the winner is settled by the blockchain platform when the winner completes all assigned computation tasks. In this manner, the efficiency of our SEAL scheme can be significantly enhanced by using the batch technique.
Besides, for failed tasks that {are} aborted by the bidder or failed to complete in time, a failed task map $fMap[i]$ is maintained on the blockchain ledger for punishment execution and financial settlement with improved robustness.

\textbf{On-chain exchange} (lines 23--40).
For each task $\Im_{l,n}$, $ l \in \mathbb{G}_i$, the winner $i$ first computes its encrypted processing result  with a symmetric key ${k_i^l}$, i.e., $\mathbf{Enc}_{k_i^l}(result_i^l)$.
Then, it generates a hash value $H_i^l=\mathbf{Hash}(k_i^l||nonce_l)$, and produces a ZKP $\pi_i^l$ using the zero-knowledge succinct noninteractive arguments of knowledge (ZK-SNARK) protocol to commit both the hash value and encrypted result. Here, $nonce_l=nonce_0+l$ is used to prevent message reply attacks. 
Next, winner $i$ sends a message
\begin{align}
ResMsg =\Big\langle\mathbf{Enc}_{pk_{\mathrm{UAV}}}(\mathbf{Enc}_{k_i^l}(result_i^l)), H_i^l,\pi_i^l \Big\rangle
\end{align}
to the UAV, and publishes $H_i^l$ on the blockchain (line 25). Upon validating the proof $\pi_i^l$, the UAV successively reveals the signed payword $h_i^l$ of $HashChain_i$ to winner $i$ (line 28) as a commitment for payment if the decryption key committed by $\pi_i^l$ is released within task deadline. Otherwise, if the validation fails, {the misbehavior of the corresponding vehicle in delivering a wrong $ResMsg$ message (which causes a failure of exchange)} is reported to the smart contract (line 30).
{Upon receiving the payword}, the vehicle {sends} its signature of ${k_i^l}'$ to the smart contract (line 32).
Only if the hash of the key matches the commitment, i.e.,
\begin{align}
\mathbf{Hash}({k_i^l}'||nonce_l) \overset{?}{=} H_i^l,
\end{align}
and the task deadline $\tau_{l,n}$ is not violated, the smart contract releases the key ${k_i^l}$ to the UAV (line 36). Otherwise, the task $l$ is added into the failed task map $fMap[i]$ of winner $i$ (line 38).
With the key ${k_i^l}$, the UAV can acquire the task results by decrypting the encrypted processing result (line 40).

\textbf{Claim} (lines 42--45). Each winning bidder $i\in \mathbb{W}$ utilizes $h_i^N$ ($1\le N\le \left|\mathbb{G}_i \right|$) received from the UAV and generates the following claim transaction to claim its due payment $p_{i}$:
\begin{align}\label{eq:claimTx}
\mathrm{tx_{claim}} = \left<\mathrm{claim},h_i^N, N+1, p_{i}, pk_{i}, T_{\mathrm{stamp}}, \sigma_{\mathrm{cla}}\right>,
\end{align}
where $\sigma_{\mathrm{cla}}$ is the signature on the hash digest of $\mathrm{tx_{claim}}$.
Based on the published root $h_i^0$, the failed task maps, and the announced auction results, the smart contract first validates $(h_i^N,N+1)$ and then computes the feasible due payment $p_{i}$ to winner $i$ if the validation succeeds, i.e.,
\begin{align}
p_{i}=\sum\nolimits_{l=1}^{N}{p_{i}^l} - \sum\nolimits_{k \in fMap[i]}{p_{i}^k}.
\end{align}
Next, it releases the due payment $p_i$ to winner $i$ from the escrow pool $esPool$ and updates UAV's deposit value.

\textbf{Refund} (lines 47--51). The smart contract closes the auction by invoking the $refund()$ to release balances to participants and update system states.

\textbf{Timeout} (lines 2, 5, 14, 18). In the smart contract, a timer is set to examine the current time and invoke corresponding functions if the time (e.g., $T_0$, $T_1$, $T_2$, $T_3$, or $T_4$) has expired.


\subsection{UAV Cost Analysis}\label{subsec:scheme2}
In its transit route, the UAV needs to schedule its computation tasks offloaded to VFC nodes for improved endurance time and service quality. The overall cost of {a} UAV includes its energy consumption and payment at location $n$, i.e.,
\begin{align}\label{eq:cost}
C_n=\varpi E_n +(1-\varpi)\lambda _p{{\sum\nolimits_{i\in \mathbb{W}}{p_{i}^{j}}}},
\end{align}
where $\varpi \in(0,1)$ is the weight factor to balance the energy cost and payment, $\lambda _p>0$ is an adjustment parameter, $\mathbb{W}$ is the set of winners of all the $J_n$ tasks, and $E_n$ is the total energy consumption of the UAV at segment $n$.

\underline{Energy consumption analysis}. The energy consumption of the UAV at each segment consists of hovering energy cost, propulsion energy cost, and transmission energy cost.
Based on \cite{9043589}, the hovering energy cost at location $n$ can be simplified as $E_{n}^{\mathrm{hov}} \!=\! P^{\mathrm{hov}} \sum\nolimits_{j=1}^{J_n}{T_{j,n}}$, where $P^{\mathrm{hov}}$ is UAV's hovering power and ${T_{j,n}}$ is the task completion time of mission $\Im_{j,n}$.
The UAV's propulsion energy cost in flying from location $n$ to location $n+1$ is $E_{n}^{\mathrm{fly}} = P_{n}^{\mathrm{fly}} L_n/V_n$, where $L_n$ is the distance between locations $n$ and $n+1$.
Finally, the transmission energy cost of mission $\Im_{j,n}$ at location $n$ is $E_{i,j}^{\mathrm{tr}} = P^{\mathrm{A2G}} T_{i,j}^{\mathrm{tr}}$, where $P^{\mathrm{A2G}}$ is UAV's transmit power.
To summarize, the overall energy cost of the UAV at segment $n$ is 
\begin{align}\label{eq:energy}
E_n &= E_{n}^{\text{fly}}+E_{n}^{\text{hov}}+\sum\nolimits_{i\in \mathbb{W}}{E_{i,j}^{\text{tr}}} \nonumber\\
&=P_{n}^{\mathrm{fly}} \frac{L_n}{V_n} + P^{\mathrm{hov}} \sum\nolimits_{j\in {\mathbb{J}_n}}{T_{j,n}} + {\sum\nolimits_{i\in \mathbb{W}}{P^{\mathrm{A2G}} T_{i,j}^{\mathrm{tr}}}}.
\end{align}

\underline{Task delay analysis}.
The task completion time of mission $\Im_{j,n}$, i.e., ${T_{j,n}}$, consists of uplink transmission time $T_{i,j}^{\mathrm{tr}}$, task processing time $T_{i,j}^{\mathrm{comp}}$, and downlink transmission time.
As the size of processing results is relatively small, the downlink G2A transmission time can be neglected \cite{9625737}.
The uplink A2G transmission delay is related to the task size ${s_{j,n}}$ and data transmission rate ${\gamma _{i}^{j}}$, so we have $T_{i,j}^{\mathrm{tr}} = {s_{j,n}}/{\gamma _{i}^{j}}$. 
Besides, according to \cite{9625737}, the task processing latency is associated with the shared computing resource $\chi _{i}^{j}$ (in CPU cycles per second), the computing intensity $\zeta_{j,n}$, and task size ${s_{j,n}}$. We have $T_{i,j}^{\mathrm{comp}}=s_{j,n}\zeta_{j,n}/\chi _{i}^{j}$. 
Therefore, the task completion time can be denoted as
\begin{align}
{T_{j,n}}={s_{j,n}}\Big(\frac{1}{\gamma _{i}^{j}} + \frac{\zeta_{j,n}}{\chi _{i}^{j}} \Big).
\end{align}

Due to the mobility of vehicles, the residual dwell time of vehicle $i\in \mathbb{I}$ in UAV's coverage should satisfy $\tau_i^R \ge {T_{j,n}}$ during task offloading.
For ease of analysis, it is assumed that the road in the UAV's coverage is straight, and the center of UAV's communication coverage lies on the central axis of the road. Then, we can obtain
\begin{align}\label{eq:dwelltime}
\tau_i^R = \frac{\Re+\varsigma_{i,n}d_{i,n}}{\overline \vartheta_{\mathrm{veh}}},
\end{align}
where vehicle $i$ is driving at a constant speed $\overline \vartheta_{\mathrm{veh}}$ in UAV's coverage {area (which is regarded as a circle with radius $\Re$ \cite{9696188})}.
In Eq. (\ref{eq:dwelltime}), $d_{i,n}$ is the horizontal distance between vehicle $i$ and the UAV at location $n$. $\varsigma_{i,n}$ is vehicle $i$'s heading direction, i.e.,
\begin{align}\label{eq:heading}
\varsigma _{i,n}=\left\{ \begin{array}{l}
	+1\text{,~if~vehicle~$i$~is~driving~towards~location~$n$;}\\
	-1\text{,~if~vehicle~$i$~is~driving~away~location~$n$.}\\
\end{array} \right.
\end{align}

\underline{Cost minimization problem}.
The optimization problem of the UAV is to minimize its operational cost under the following practical constraints ($\forall i\!\in\! \mathbb{I},\forall j\in \mathbb{J}_n$)\footnotemark[4].\footnotetext[4]{{In the future, we will further investigate UAV's global energy consumption minimization problem across $N$ sensing locations under battery supply limits. As the energy consumption at each location affects the energy budget at subsequent locations, an energy deficit (indicating the energy consumption deviation from the average energy budget) \cite{{9043589}} can be further considered to break the energy supply linkage across $N$ locations, thereby facilitating the optimization process.}}
\begin{align}\label{eq:optimization}
&\min_{\overrightarrow{\beta},\overrightarrow{{p}}}\varpi\left\{{\sum_{j\in {\mathbb{J}_n}}{\sum_{i\in \mathbb{I}}{\!\left(\! \frac{P^{\text{hov}}\zeta _{j,n}}{\chi _{i}^{j}}\!+\!\frac{P^{\text{A2G}}\!+\!P^{\text{hov}}}{\gamma _{i}^{j}} \!\right)\! }}}s_{j,n}\beta _{i}^{j} \right. \nonumber \\
&~~~\left.+P_{n}^{\mathrm{fly}} L_n/V_n\right\} +(1-\varpi)\lambda _p{\sum\nolimits_{j\in {\mathbb{J}_n}}{\sum\nolimits_{i\in \mathbb{I}}{p_{i}^{j}}}}\beta _{i}^{j}
\end{align}
\begin{numcases}{s.t.}	
{{T_{j,n}}}\beta _{i}^{j}\leq \min\{\tau _{j,n},\tau _{i}^{R}\},  \label{eq:cons1} \hfill \\
\pi( \chi _{i}^{j},b_{i}^{j} ) \ge 0, \label{eq:cons5} \hfill \\
\pi({\overrightarrow{\mathcal{B}}_{i}^{j}}^*,\overrightarrow{\mathcal{B}}_{\!-\!i}^j)\!\geq\! \pi(\overrightarrow{\mathcal{B}}',\overrightarrow{\mathcal{B}}_{\!-\!i}^j),\forall \overrightarrow{\mathcal{B}}' \!\ne\! {\overrightarrow{\mathcal{B}}_{i}^{j}}^*, \label{eq:cons6} \hfill \\
\sum\nolimits_{i\in \mathbb{I}}{\beta _{i}^{j}}=1, \label{eq:cons7} \hfill \\
\beta _{i}^{j}\in \{0,1\},{p_{i}^{j}}\ge 0. \label{eq:cons8} \hfill
\end{numcases}

The decision variables in the above problem are $\overrightarrow{\beta}$ and $\overrightarrow{{p}}$.
Constraint (\ref{eq:cons1}) is the task deadline constraint of vehicles, indicating that the task completion time ${{T_{j,n}}}$ should be less than both the task completion deadline $\tau _{j,n}$ and the vehicular residual dwell time $\tau _{i}^{R}$.
Constraints (\ref{eq:cons5}) and (\ref{eq:cons6}) are IR and CIC constraints for each vehicle, respectively.
Constraint (\ref{eq:cons7}) indicates that each task can be assigned to at most one vehicle. 

\subsection{Multi-Round SRC Auction Design}\label{subsec:scheme3}
Note that the relaxed problem in formula (\ref{eq:optimization}) with fixed payment $\overrightarrow{p}$ and constraints (\ref{eq:cons7})--(\ref{eq:cons8}) is a typical weighted set cover problem and is NP-hard. Thereby, the problem in formula (\ref{eq:optimization}) is NP-hard, and it is nontrivial to attain the optimal solution in polynomial time. 
Besides, conventional single-parameter truthful auctions \cite{7118253,8493354,8675981,9354536,8489932,8169049} cannot be directly applied, as our SRC auction is a double-parameter truthful auction with both bidding price and computation resource supply truthfulness for heterogeneous tasks. 
Next, we introduce the sufficient and necessary conditions for satisfying CIC in SRC auctions in the following theorem.
\\
\textbf{Theorem 1}. \emph{A SRC auction mechanism is combinatorial incentive-compatible if the following two properties hold:}
\begin{itemize}
  \item Monotonicity. \emph{For each task $\Im_{j,n}$, given that other bidders' strategies are fixed, any bidder $i\in \mathbb{I}$ wins the auction with bid $({\chi_{i}^{j}},{b_{i}^{j}})$ still wins by bidding $({{\chi_{i}^{j}}'},{{b_{i}^{j}}'})$  with ${{\chi_{i}^{j}}'}>{\chi_{i}^{j}}$ and ${{b_{i}^{j}}'}<{b_{i}^{j}}$.}
  \item Critical payment. \emph{Any winner $i$ with bid $({\chi_{i}^{j}},{b_{i}^{j}})$ of task $\Im_{j,n}$ is paid the critical payment, i.e., the supremum of all bidding prices ${b_{i}^{j}}$'s such that $({\chi_{i}^{j}},{{b_{i}^{j}}'})$ still wins, i.e., $p_{i*,j}^{\mathrm{CP}}=\mathrm{sup}\{{b_{i}^{j}}' |{\beta_{i}^{j}}'=1\}$, given the bids of others remain unchanged.}
\end{itemize}
\begin{IEEEproof}
A similar proof can refer to \cite{NISAN2015477} (in Sect. 5.4).
We move the detailed proof to Appendix~A in the supplementary material due to the page limitation.
\end{IEEEproof}

Utilizing the rationale provided in Theorem 1, we design a multi-round SRC auction mechanism for approximate UAV's cost minimization with strategy proofness and computational efficiency.
Algorithm~\ref{Algorithm2} summarizes the auction workflow, which includes three consecutive phases: candidate group formulation, optimal worker selection, and payment determination.

\begin{algorithm}[!t]\begin{small}
   \caption{Task Offloading Scheduling with Winner Selection and Pricing in Multi-round SRC Auction}\label{Algorithm2}
    \begin{algorithmic}[1]
        \STATE \textbf{Input: }$\mathbb{I}$, $\mathbb{{J}}_n$, $\Im_{j,n}$, $\mathcal{B}_i$, $\Re$, $\varsigma_{i,n}$, $d_{i,n}$, $\hbar_n$, $\overline \vartheta_{\mathrm{veh}}$, $\overline{\chi _i}$, $P^{\mathrm{A2G}}$, $P^{\text{hov}}$;
        \STATE \textbf{Output: }$\overrightarrow{\beta }$, $\overrightarrow{p^{\mathrm{CP}}}$;
        \STATE \textbf{Initialize: }$\beta_{i}^{j}\leftarrow 0$, $p_{i}^{j}\leftarrow 0$, $\Gamma _i \leftarrow \emptyset$, $\mathbb{W} \leftarrow \emptyset$, $\mathbb{G}_{i}\leftarrow \emptyset$;
        \STATE Sort all tasks in set $\mathbb{J}_n$ in decreasing order of the urgency degree and obtain a new task set $\mathbb{J}_n'$;
        \FOR {$i =1,2,\cdots, \left|\mathbb{I}\right| $}
            \STATE Obtain the feasible task set $\Gamma _i$ using Eq. (\ref{eq:taskset});
        \ENDFOR
        \FOR {$j =1,2,\cdots, \left|\mathbb{J}_n'\right| $}
            \STATE \textcolor{gray}{\#\,Select feasible workers as candidates.\,\#}
            \STATE Obtain the feasible candidate set $\mathbb{C}_{j,n}$ using Eq. (\ref{eq:candidateset});
            \FOR {$i  \in \mathbb{C}_{j,n} $}
                \STATE Compute the marginal cost factor $\digamma(\chi _{i}^{j},b_{i}^{j})$ using Eq. (\ref{eq:MCF});
            \ENDFOR
            \STATE \textcolor{gray}{\#\,Find the bidder with the minimum marginal cost factor.\,\#}
            \STATE $i*=\text{arg}\min_{i} \left\{ \digamma (\chi _{i}^{j},b_{i}^{j}): i \in \mathbb{C}_{j,n} \right\}$;
            \STATE $\beta _{i*,j,n}\leftarrow 1$, $\mathbb{W} \leftarrow \mathbb{W}\cup \{i*\}$;
            \STATE $\mathbb{G}_{i*}\leftarrow \mathbb{G}_{i*} \cup \{j\}$;
            \STATE $\Gamma _{i} \leftarrow \Gamma _{i} \setminus \{j\}, \forall i \in \mathbb{C}_{j,n}$;
            \STATE \textcolor{gray}{\#\,Calculate residual computing resource.\,\#}
            \STATE $\overline{\chi_{i*}} \leftarrow \overline{\chi_{i*}} - \chi_{i*}^j$;
        \ENDFOR
        \FOR {$j =1,2,\cdots, \left|\mathbb{J}_n'\right| $}
            \STATE $\mathbb{C}_{j,n}'\leftarrow \mathbb{C}_{j,n}\backslash \{i*\}$;
            \STATE Perform winner selection process in lines $11$--$15$ with input $\mathbb{C}_{j,n}'$ and compute a new winner $k$;
            \STATE Compute the virtual bidding price $\widetilde{{{b}}}_{i*,j}$ using Eq. (\ref{eq:virtualbidding});
            \STATE \textcolor{gray}{\#\,Calculate critical payment.\,\#}
            \STATE $p_{i*,j}^{\mathrm{CP}} \leftarrow \widetilde{{{b}}}_{i*,j}$;
        \ENDFOR
    \end{algorithmic}\end{small}
\end{algorithm}

\underline{1) Candidate group formulation} (lines 4--10).
In the task area, due to the high mobility of vehicles and UAVs and the heterogeneity of tasks in terms of task delay and resource demand, part of vehicles may not complete the assigned tasks in time or lack sufficient computation resource for task processing, causing a failure of task offloading. Therefore, efficient task allocation is needed with consideration of the heterogeneity of both tasks and vehicles.
Generally, the task with a higher urgency degree needs to be offloaded with a higher priority. By sorting the urgency degree of all tasks in set $\mathbb{J}_n$ in decreasing order, a new task set is obtained as $\mathbb{J}_n'$ (line 4).
\\
\textbf{Definition 5 (Feasible Task Set).} \emph{For each vehicle $i \in \mathbb{I}$, its feasible task set $\Gamma _i$ can be formed by sequentially adding every feasible task from $\mathbb{J}_n'$ (line 6) as below:
\begin{align}\label{eq:taskset}
\Gamma _i=&\left\{ j\Big| {s_{j,n}}\Big(\frac{1}{{\gamma _{i}^{j}}} + \frac{\zeta_{j,n}}{\chi _{i}^{j}} \Big)\leq \min\{\tau _{j,n},\tau _{i}^{R}\}\right.\nonumber\\[-3pt]
&~~~~~~~~\mathrm{and}~\sum\nolimits_{j\in \Gamma _i}{\chi _{i}^{j}}\leq \overline{{\chi _{i}}},\,\forall \Big. j\in \mathbb{J}_n' \Big\},
\end{align}
where $\overline{{\chi _{i}}}$ is the available computing resource of vehicle $i$.}
\\
\textbf{Definition 6 (Feasible Candidate Set).} \emph{For each task $\Im_{j,n}$, the vehicles that satisfy constraint (\ref{eq:cons1}) can be included into the feasible candidate set $\mathbb{C}_{j,n}$ (line 10), which is defined as:
\begin{align}\label{eq:candidateset}\resizebox{0.90\hsize}{!}{$
\mathbb{C}_{j,n}\!=\!\left\{ i\Big| {s_{j,n}}\Big(\frac{1}{{\gamma _{i}^{j}}} \!+\! \frac{\zeta_{j,n}}{\chi _{i}^{j}} \Big)\!\leq\! \min\{\tau _{j,n},\tau _{i}^{R}\}, \forall j\!\in\! \Gamma _i,i\!\in\! \mathbb{I} \right\}\!.\!$}
\end{align}}

\underline{2) Optimal worker selection} (lines 11--20).
For every task in $\mathbb{J}_n'$, the winner to perform the task is determined based on the \emph{marginal cost factor (MCF)}, which indicates the marginal cost increment (including the energy cost and payment cost) of bidder~$i$ to the UAV in the task (line 12).
\\
\textbf{Definition 7 (Marginal Cost Factor).} \emph{For every candidate in $\mathbb{C}_{j,n}$, its MCF $\digamma(\chi _{i}^{j},b_{i}^{j})$ in performing task $\Im_{j,n}$ is defined as:
\begin{align}\label{eq:MCF}\resizebox{0.894\hsize}{!}{$
\digamma(\chi _{i}^{j},b_{i}^{j}) \!=\! \varpi s_{j,n} \Big( \frac{P^{\text{hov}}\zeta _{j,n}}{\chi _{i}^{j}}\!+\!\frac{P^{\text{A2G}}+P^{\text{hov}}}{\gamma _{i}^{j}} \Big) \!+\!(1\!-\!\varpi)\lambda _pb_{i}^{j}.\!$}
\end{align}}

In every iteration, the UAV calculates the optimal candidate $i*$ for task $\Im_{j,n}$ that incurs the lowest MCF over $\mathbb{C}_{j,n}$ as the winner (line 15), i.e.,
\begin{align}\label{eq:winner}
\beta _{i*}^j=\left\{ \begin{array}{l}
	1,\,if\ i*=\text{arg}\min_{i} \left\{ \digamma (\chi _{i}^{j},b_{i}^{j}): i \!\in\! \mathbb{C}_{j,n} \right\};\\[-2pt]
	0,\,otherwise.\\[-5pt]
\end{array} \right.
\end{align}
In our work, the auction for each task in the set $\mathbb{J}_n'$ is carried in a sequential manner, where the tasks with higher urgency degrees are executed earlier.
After the auction processes of all tasks in $\mathbb{J}_n'$ finish, the set of winners (line 16) can be derived as
\begin{align}\label{eq:winnerset}
\mathbb{W}=\left\{ i\Big|\beta_{i}^{j}=1, j \in \mathbb{J}_n'\right\}.
\end{align}
Besides, the set of allocated tasks to winner $i*\in \mathbb{W}$ (line 17) is 
\begin{align}\label{eq:winnertasks}
\mathbb{G}_{i*}=\left\{ j\Big|\beta _{i*}^j=1, j \in \mathbb{J}_n'\right\},
\end{align}
where $\mathbb{G}_{i}\cap \mathbb{G}_{i'}=\varnothing$, $\forall i \ne i'$.

\underline{3) Payment determination} (lines 22--28).
After determining the winner of each task, we develop a pricing method by employing the critical payment defined in Theorem 1 to decide the payment to each winner with combinatorial truthfulness guarantees.
\\
\textbf{Definition 8 (Critical Bidder).} \emph{The critical bidder is defined as the virtual winner $k$ that wins the auction of task $\Im_{j,n}$ when excluding the original winner $i*$ from the candidate set, i.e., $k = \text{arg}\min_{i} \big\{ \digamma (\chi _{i}^{j},b_{i}^{j}): i \!\in\! \mathbb{C}_{j,n}\backslash\{i*\} \big\}$.}

Particularly, for each mission $\Im_{j,n}$, a new winner selection procedure is executed over all candidates in $\mathbb{C}_{j,n}$ except the winner $i*$ (line 23). Then, a new winner $k \in \mathbb{C}_{j,n}\backslash\{i*\}$ (i.e., critical bidder) can be chosen (line 24). 
Let $\widetilde{{{b}}}_{i*,j}$ be the \emph{virtual bidding price} of bidder $i*$, which is defined as its maximum bidding price that substitutes bidder $k$ as the winner (line 25). It indicates that \begin{align}
\digamma(\chi _{k}^j,b_{k}^j) = \digamma(\chi _{i*}^j,\widetilde{{{b}}}_{i*,j}).
\end{align}
By solving the above equation, $\widetilde{{{b}}}_{i*,j}$ can be derived as:
\begin{align}\label{eq:virtualbidding}
\widetilde{{{b}}}_{i*,j}=&\frac{\varpi}{(1-\varpi)\lambda _p} s_{j,n}\Big[ P^{\text{hov}}\zeta _{j,n}\big(\frac{1}{\chi _{k}^j}-\frac{1}{\chi _{i*}^j}\big) + \Big. \nonumber \\
&\Big. \big( P^{\text{A2G}}+P^{\text{hov}} \big) \big( \frac{1}{\gamma _{k}^j}-\frac{1}{\gamma _{i*}^j} \big) \Big] +b_{k}^j.
\end{align}
Then, the critical payment $p_{i*,j}^{\mathrm{CP}}$ is derived for every winner $i*$ by setting its value equal to the virtual bidding price $\widetilde{{{b}}}_{i*,j}$ (line 27).

\emph{\textbf{Remark}.} For each task to be offloaded, we iteratively execute the SRC auction mechanism to obtain the optimal winners and their corresponding payments in each auction round.

\subsection{Theoretical Analysis}\label{subsec:analysis}
In this subsection, we first show that SEAL satisfies CIC (Lemma 1) and IR (Lemma 2).
Based on these two lemmas, we then prove its combinatorial strategy-proofness in Theorem 2. Next, we prove the desired properties of SEAL including fairness and privacy protection in Theorems 3 and 4, respectively. Finally, we analyze the complexities of SEAL in Theorem 5. 

\emph{\textbf{Lemma 1}. SEAL satisfies combinatorial incentive compatibility (CIC) for both computation resource supply and bidding price.}
\begin{IEEEproof}
Please refer to Appendix~B. 
\end{IEEEproof}

\emph{\textbf{Lemma 2}. SEAL satisfies individual rationality (IR).}
\begin{IEEEproof}
Please refer to Appendix~C. 
\end{IEEEproof}

\emph{\textbf{Theorem 2}. SEAL is a combinatorial strategy-proof auction mechanism.}
\begin{IEEEproof}
According to Lemmas 1 and 2, both CIC and IR properties are satisfied. 
Based on Definition 1, our SRC auction mechanism ensures combinatorial strategy-proofness.
Besides, numerical results in Fig.~\ref{fig:simu5} also validate it.
\end{IEEEproof}

\emph{\textbf{Theorem 3}. SEAL can preserve participants' bidding privacy.}
\begin{IEEEproof}
{Please refer to Appendix~D.} 
\end{IEEEproof}

\emph{\textbf{Theorem 4}. SEAL is a fair auction mechanism, i.e., it guarantees both participation fairness and exchange fairness.}
\begin{IEEEproof}
{Please refer to Appendix~E.} 
\end{IEEEproof}

\emph{\textbf{Theorem 5}. The computational complexity and communication complexity of SEAL are $\mathcal{O}(J_n I \log(I))$ and $\mathcal{O}(I (M \cdot bit_{\pi}+ bit_{c}))$, respectively.}

\begin{IEEEproof}
{We move the detailed proof to Appendix~F, as well as summarize the computation and communication complexities in each phase of SEAL in Table~\ref{tableComplexity}.} 
\end{IEEEproof}

\begin{table}[]\setlength{\abovecaptionskip}{-0.05cm}
\caption{Computation and Communication Complexity in SEAL}\label{tableComplexity}
\resizebox{1.01\linewidth}{!}{
\begin{tabular}{lcc}\hline
                                                             & \textbf{Comp. Complexity}                                                                                           & \textbf{Comm. Complexity}                                                            \\ \hline
Init                                                         & $\mathcal{O}(I)$                      & $\mathcal{O}(I\cdot bit_c)$                  \\
Deposit                                                      & $\mathcal{O}(I)$                                                                                                    & $\mathcal{O}(I\cdot M)$                                                              \\
\begin{tabular}[c]{@{}l@{}}Off-chain auction\end{tabular} & $\mathcal{O}(\sum\nolimits_{j\!=\!1}^{J_n}{I_j} \!+\! J_n\left|\mathbb{C}_{j,n}\right|\log(\left|\mathbb{C}_{j,n}\right|))$ & $\mathcal{O}(M\cdot bit_p)$ \\
Commit                                                       & $\mathcal{O}(\sum\nolimits_{i=1}^{\left|\mathbb{W}\right|}{\left|\mathbb{G}_{i}\right|})$                           & $\mathcal{O}(M)$                                                              \\
\begin{tabular}[c]{@{}l@{}}On-chain exchange\end{tabular} & $\mathcal{O}(\sum\nolimits_{i=1}^{\left|\mathbb{W}\right|}{\left|\mathbb{G}_{i}\right|})$                           & $\mathcal{O}(M\left|\mathbb{W}\right|bit_{\pi})$                                        \\
Claim                                                        & $\mathcal{O}(\sum\nolimits_{i=1}^{\left|\mathbb{W}\right|}{\left|\mathbb{G}_{i}\right|})$                           & $\mathcal{O}(M\left|\mathbb{W}\right|)$                                        \\ [0.2pt]
Refund                                                       & $\mathcal{O}(I)$                                                                                                    & $\mathcal{O}(I\cdot M)$                                                              \\ \hline
\textbf{SEAL}                                                & $\mathcal{O}(J_n I \log(I))$                                                                                        & $\mathcal{O}(I (M \!\cdot\! bit_{\pi}\!+\! bit_{c}))$ \\ \hline
\end{tabular}}\vspace{-0.0cm}
\end{table}

\section{PERFORMANCE EVALUATION}\label{sec:SIMULATION}

\begin{table}[!t]\setlength{\abovecaptionskip}{-0.0cm}
    \begin{center}
        \caption{Simulation Parameters}\label{table2}
        \begin{tabular}{cc|cc|cc}
        \hline 
        {Param} & {Value} & {Param} & {Value} & {Param} & {Value}  \\ \hline
            $N$ &$30$ &$K$ &$1000$ &$J_n$ &$[100,300]$ \\
            $\hbar_n$ &$50$m & $s_{j,n}$ &$[3,9]$ Mb & $\tau_{j,n}$ &$[1.0,2.5]$s \\
            $\overline{{\chi _{i}}}$ &$[0.5,2.0]$GC/s &$\varpi$&$0.5$ &$\zeta_{j,n}$ &$50$ C/Mb \\
            $\varphi_{j,n}$ &$[0.1,1]$ &$\Re$&$250$m &$L_n$&$500$m \\
            $\gamma _{i ,n}$&$6$ Mbps &$V_{\min}$ &$2$ m/s &$V_{\max}$&$20$ m/s \\
            ${\lambda_p}$&$40$& ${\vartheta_{\mathrm{veh}}^{\min }}$&$30$ km/h &${\vartheta_{\mathrm{veh}}^{\max }}$ &$80$ km/h\\[2pt]
            $\phi_i$&$[1,9]$ &$P^{\mathrm{A2G}}$&$0.2$W &$P^{\mathrm{hov}}$&$500$W \\
            \hline 
        \end{tabular}
    \end{center}\vspace{-0.32cm}
\end{table}
\subsection{Simulation Settings}\label{subsec:evalution1}
We conduct simulations on a real-world data set from the mobility traces of taxi cabs in San Francisco \cite{Traceset2009}, which contains GPS coordinates of about $500$ taxis gathered over a month in the San Francisco Bay Area. 
There are $30$ sensing locations uniformly distributed in the area. A UAV flies at a fixed altitude $50$m and sequentially visits each location via a straight-line trajectory.
The number of computation missions produced at each location is randomly selected between $100$ and $300$. The task size and task deadline follow the uniform distribution within $[3,9]$Mb and $[1.0,2.5]$s, respectively. 
The computation intensity is $\zeta\!=\!50$ CPU cycles/Mb. The idle computing resource of ground vehicles is randomly selected from $[0.5,2.0]$ GC/s (GC $=10^9$ CPU cycles) \cite{8672604}. 
We define $\Theta({\chi_{i}^{j}})=\phi_i {\chi_{i}^{j}}+c_0$ as the private cost valuation of bidder $i$ \cite{8489932}, where $\phi_i$ is its unit cost of computing resource and $c_0$ is the fixed cost.


The Intel SGX SDK\footnotemark[5] is adopted to implement the SRC auction algorithm, where the SGX enclave serves as the auctioneer and runs the auction algorithm. 
The software attestation process is implemented where bidders verify whether the auction program is correctly coded and loaded in the enclave. After attestation, bidders send their encrypted combinatorial bids to the SGX enclave which loads the ciphertexts via the \emph{ecall} function. 
We implement the smart contracts in JavaScript in a local simulated environment using Hyperledger Caliper\footnotemark[6], which is a widely used customizable benchmarking tool for blockchains such as Hyperledger Fabric and Ethereum.
A Caliper adaptor is programmed via Fabric Client SDK using Node.js to interact with the blockchain platform.
As the reference implementation, 3 ordering service nodes (OSNs) run the Apache Kafka protocol to reach consensus on generated transactions. 
Besides, the ZKP is realized based on the open source library libsnark\footnotemark[7], and the Keccak-256 is adopted as the one-way hash function for efficient hashchain creation.
The simulation parameters are summarized in Table \ref{table2} \cite{9043589,8672604}.
\footnotetext[5]{https://software.intel.com/content/www/us/en/develop/topics/software-guard-extensions/sdk.html}
\footnotetext[6]{https://hyperledger.github.io/caliper/}
\footnotetext[7]{https://github.com/scipr-lab/libsnark}


\subsection{System Overheads}\label{subsec:evalution2}
We compare the SEAL with the following representative privacy-preserving auction schemes in terms of system overhead. 
\begin{itemize}
  \item \emph{ARMOR scheme \cite{8493354}:} it utilizes cryptographic tools including HE and garbled circuits to preserve users' privacy in combinatorial spectrum auctions.
  \item \emph{BidGuard scheme \cite{8169049}:} it leverages the exponential mechanism in DP for bid perturbation to prevent bid privacy inference in truthful MCS auctions.
  \item \emph{SAFE scheme \cite{9296813}:} it leverages TEE for bid privacy protection in general single-round auctions on smart contract systems, where the batch payment and double-parameter truthfulness in multi-round auctions are not supported.
\end{itemize}
Note that these schemes focus on distinct auction formats in different applications. To be objective and fair, we implement the multi-round SRC auction in the above three schemes under UAV computation offloading scenarios, and other operations follow the original schemes.

\begin{table}[!t]\setlength{\abovecaptionskip}{-0.06cm}
\caption{Computation and communication overheads in SEAL under small-scale and large-scale auctions}\label{tabCompComm}
\resizebox{1.01\linewidth}{!}{
\begin{tabular}{l|cccc}
\hline
                  & \multicolumn{2}{c}{$J=10$,~$I=10$} & \multicolumn{2}{c}{$J=100$,~$I=50$} \\ \cline{2-5}
                  & Comp.(ms)    & Comm.(KB)    & Comp.(ms)     & Comm.(KB)    \\ \hline
Init              & 2.6           & 14.36         & 2.6            & 67.01         \\
Off-chain aution  & 2.8           & 0.15          & 45             & 2.42          \\
Commit            & 1.3           & 7.85          & 7.1            & 32.53         \\
On-chain exchange & 61            & 9.5           & 133            & 93.89         \\
Claim             & 29            & 1.59          & 29             & 7.97          \\ \hline
Total             & 96.7          & 33.45         & 216.7          & 203.82        \\ \hline
\end{tabular}}\vspace{-0.1cm}
\end{table}

\begin{table}[!t]\setlength{\abovecaptionskip}{-0.32cm}
\caption{Comparison of computation overheads in four schemes ($J=100$)}\label{tabComp}
\begin{center}
\begin{tabular}{l|ccccc}\hline
& \multicolumn{5}{c}{Computation overhead (ms)}         \\ \cline{2-6}
          & $I=10$      & $I=20$      & $I=30$      & $I=40$      & $I=50$      \\ \hline
ARMOR    & 2.6$\times 10^3$    & 9.7$\times 10^3$    & 23.3$\times 10^3$   & 35.2$\times 10^3$   & 48.8$\times 10^3$   \\
BidGuard & 13.6  & 24.5  & 37.4  & 49.8  & 62.5  \\
SAFE     & 188.2 & 199.4   & 210.1 & 229.1 & 357.8 \\
SEAL     & 100.1 & 121.6 & 147.0   & 177.5 & 216.7 \\ \hline
\end{tabular}
\end{center}\vspace{-0.28cm}
\end{table}

\begin{table}[!t]\setlength{\abovecaptionskip}{-0.5cm}
\caption{Comparison of communication overheads in four schemes ($J=100$)}\label{tabComm}
\begin{center}
\resizebox{1.0\linewidth}{!}{
\begin{tabular}{l|ccccc}\hline
& \multicolumn{5}{c}{Communication overhead (KB)}         \\ \cline{2-6}
         & $I=10$      & $I=20$      & $I=30$      & $I=40$      & $I=50$      \\ \hline
ARMOR    & 2.1$\times 10^3$   & 4.7$\times 10^3$   & 10.2$\times 10^3$   & 15.8$\times 10^3$   & 21.3$\times 10^3$   \\
BidGuard & 19.75 & 36.05 & 53.35  & 69.66  & 85.96  \\
SAFE     & 104.3 & 132.8 & 178.2  & 235.8  & 292.3  \\
SEAL     & 36.49 & 68.99 & 101.08 & 133.17 & 165.25 \\ \hline
\end{tabular}}
\end{center}\vspace{-0.28cm}
\end{table}

\begin{table}[!t]\setlength{\abovecaptionskip}{-0.25cm}
\caption{Communication complexity in commit-then-commit operations in SEAL and SAFE schemes under high-frequency trading}\label{tabComComplexity}
\begin{center}
\begin{tabular}{lc}
\hline
     & \begin{tabular}[c]{@{}c@{}}Comm. complexity in commit-then-commit operations\\ under high-frequency payment ($\left|\mathbb{W}\right| \!<\!< J$)\end{tabular} \\ \hline
SEAL & $\mathcal{O}(M\cdot \left|\mathbb{W}\right|\cdot bit_{\pi})\thickapprox \mathcal{O}(M\cdot bit_{\pi})$                                                                                                                        \\
SAFE & $\mathcal{O}(M\cdot J\cdot bit_{\pi})$                                                                                                                       \\ \hline
\end{tabular}
\end{center}\vspace{-0.3cm}
\end{table}



\textbf{1) Computation \& communication overheads.}
Table~\ref{tabCompComm} shows the computation and communication overheads in each auction phase of SEAL for small-scale and large-scale auctions. 
It can be seen that the execution time and communication cost of SEAL are very low under both small-scale and large-scale scenarios, as the combinatorial bids are processed in plaintext inside the trusted enclave. The on-chain exchange phase occupies a majority of the time and communication cost owing to the creation and verification of ZKP for task result/payment delivery.

\begin{figure}[!tbp]\setlength{\abovecaptionskip}{-0.05cm}
\begin{minipage}[t]{0.243\textwidth}
\centering
    \includegraphics[height=3.3cm,width=\linewidth]{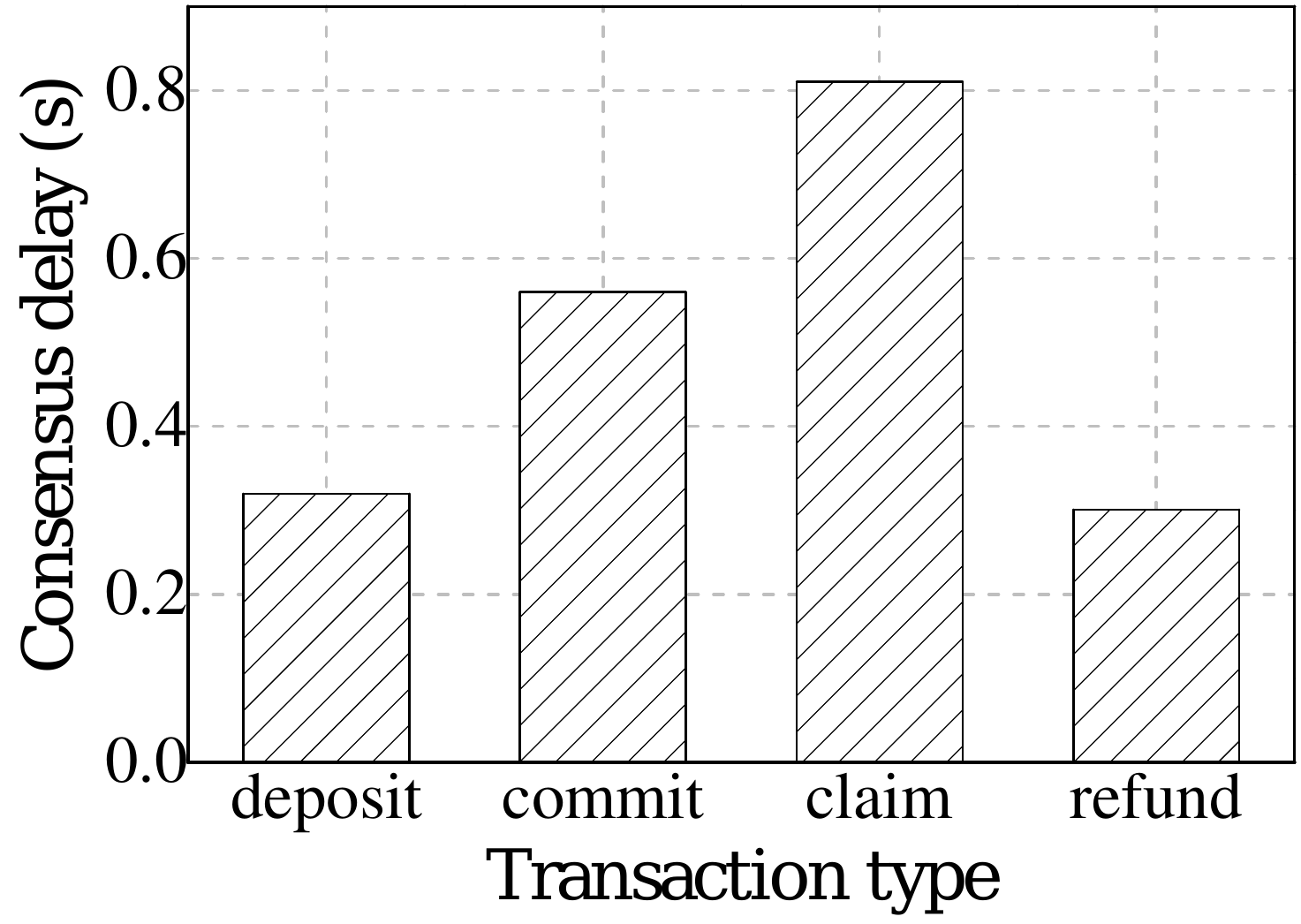}
    \caption{Consensus delay for different transaction types in the smart contract.}\label{fig:simu12}
\end{minipage}~~
\begin{minipage}[t]{0.248\textwidth}
\centering
    \includegraphics[height=3.4cm,width=\linewidth]{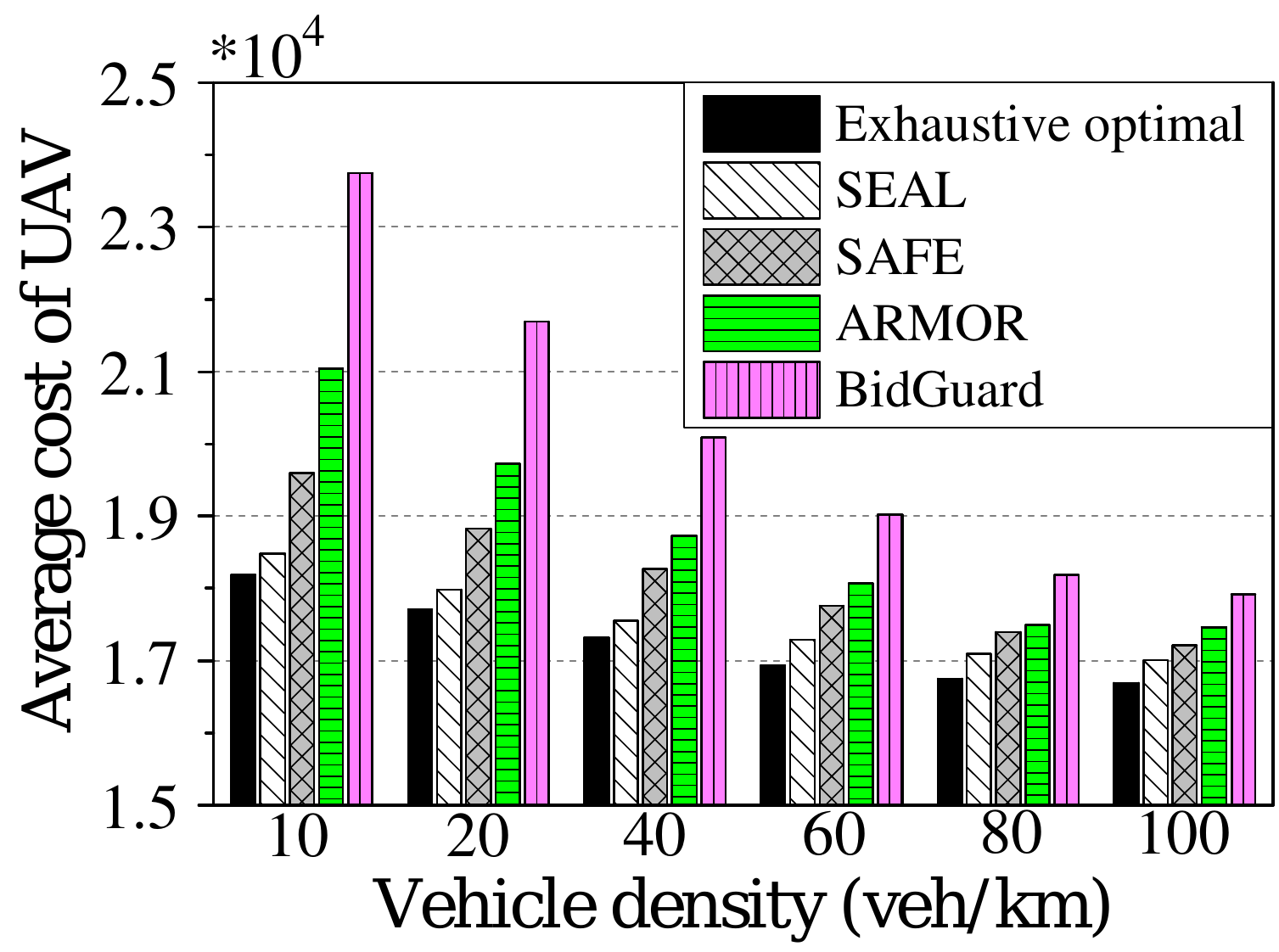}
    \caption{{Average cost of UAV vs. vehicle density in five schemes.}}\label{fig:simu2}
\end{minipage}
\end{figure}

Then we compare the computation and communication overheads of SEAL with other three schemes in Tables~\ref{tabComp} and \ref{tabComm}, respectively. Obviously, BidGuard has the smallest system overheads as it only sends perturbed bids to the auctioneer for winner and payment determination. Meanwhile, a large bid utility decrease may occur in BidGuard for practical use, especially requiring strong privacy provisions (as shown in Fig.~\ref{fig:simu2}). Besides, SEAL outperforms ARMOR and SAFE in attaining smaller computation and communication overheads given different number of bidders. For example, when $I=30$, SEAL needs 147ms with about 101KB communication cost, SAFE requires about 210ms with near 178KB communication overheads, while ARMOR spends over 23s with about 10MB communication costs.

\begin{figure}[!tbp]\setlength{\abovecaptionskip}{-0.05cm}
\begin{minipage}[t]{0.245\textwidth}
\centering
    \includegraphics[height=3.37cm,width=\linewidth]{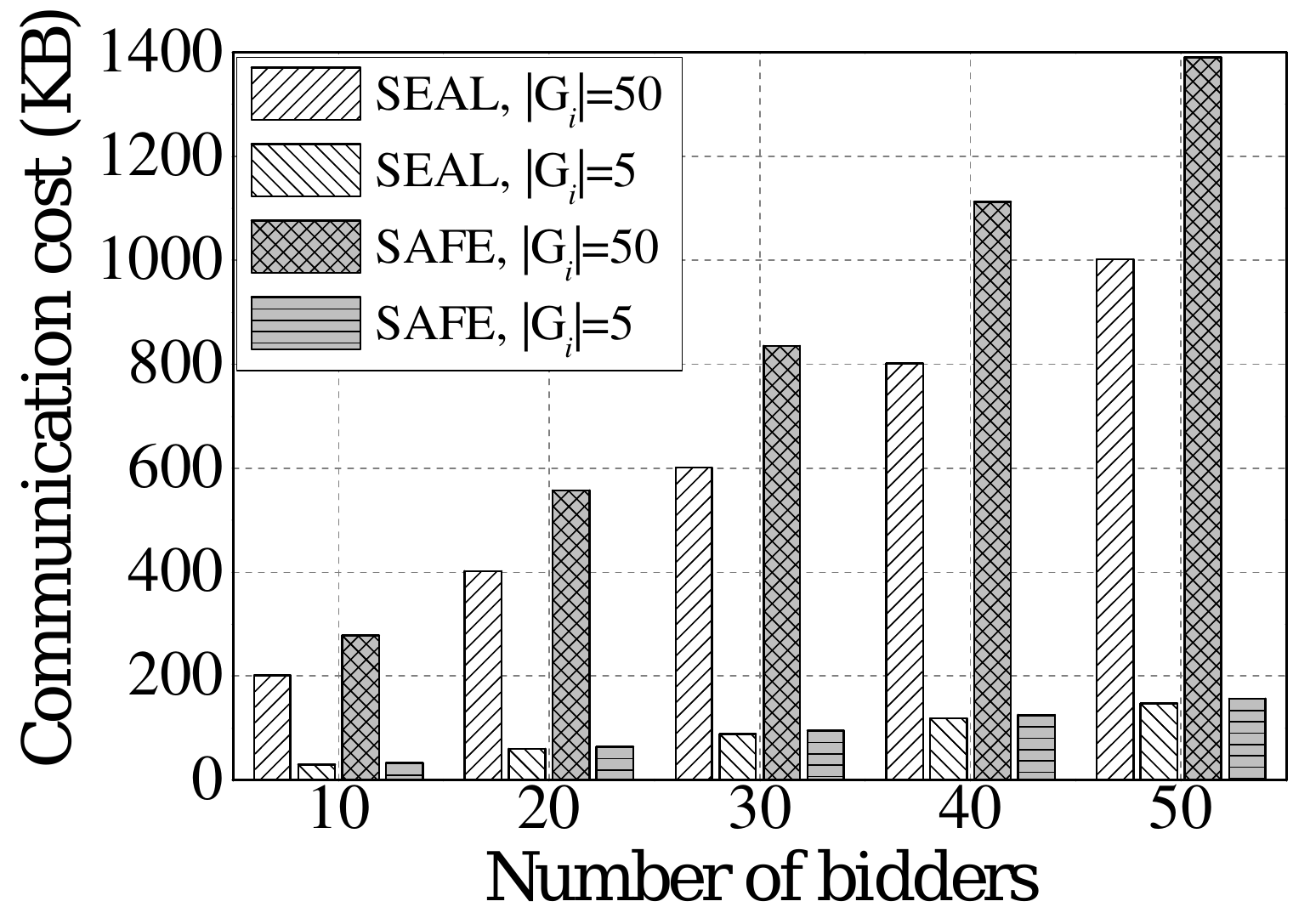}
    \caption{Communication cost of SEAL and SAFE schemes in commit-then-commit operations under low-frequency and high-frequency payment scenarios.}\label{fig:simu10}
\end{minipage}~~
\begin{minipage}[t]{0.245\textwidth}
\centering
    \includegraphics[height=3.37cm,width=\linewidth]{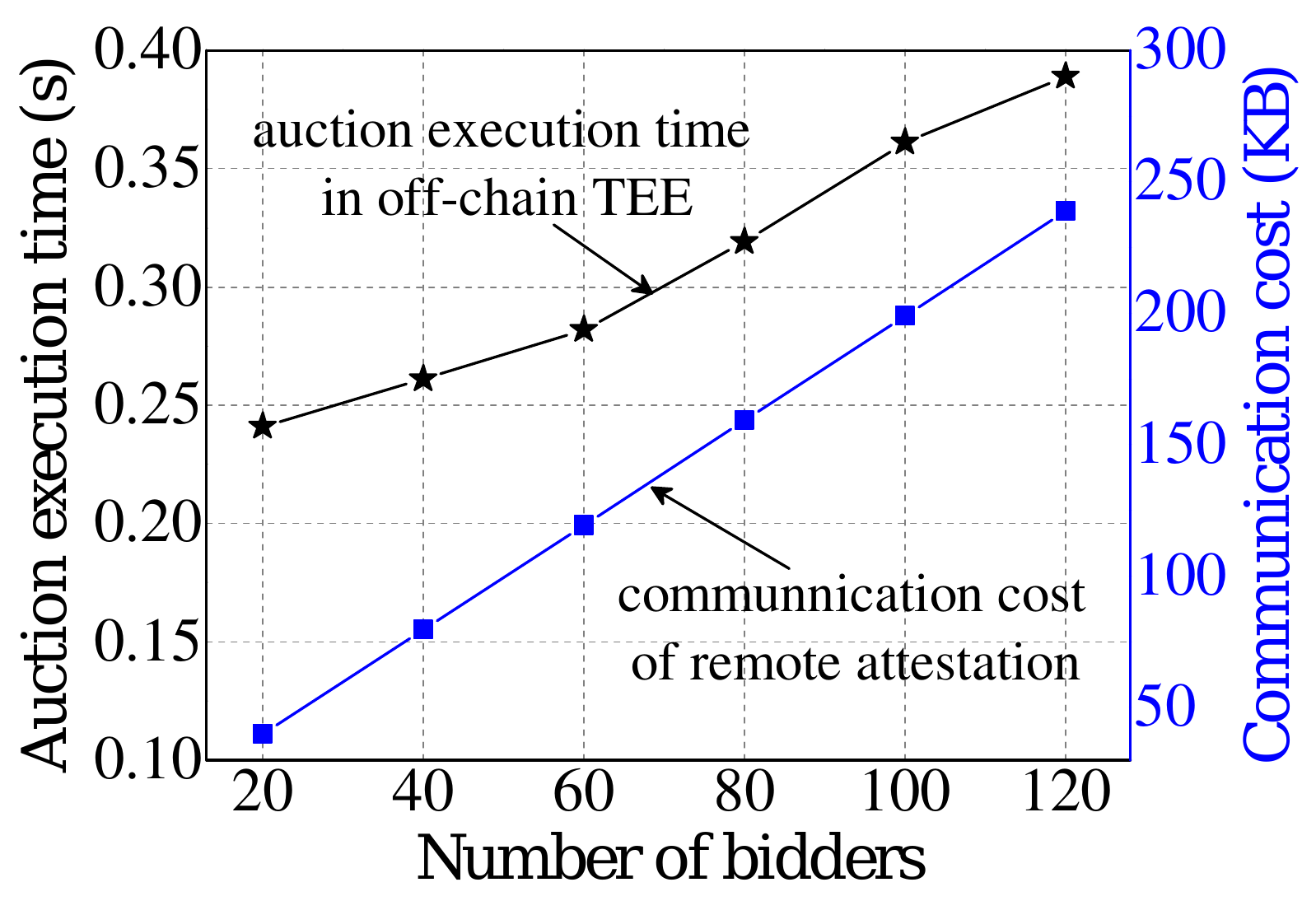}
    \caption{Off-chain auction execution time in TEE and communication cost of remote attestations with different number of bidders.}\label{fig:simu9}
\end{minipage}
\end{figure}

Next, we show the performance of hashchain micropayment in SEAL. As seen in Table~\ref{tabComComplexity}, under high-frequent payment scenarios, SEAL enjoys a much smaller communication complexity (i.e., $\mathcal{O}(M\cdot bit_{\pi})$) than that in SAFE (i.e., $\mathcal{O}(M\cdot J\cdot bit_{\pi})$) in commit-then-commit operations (i.e., the commit, on-chain exchange, and claim phases).
Besides, Fig.~\ref{fig:simu10} compares the communication cost with the SAFE scheme in commit-then-commit operations in multi-round SRC auctions, under both low-frequency ($\left| \mathbb{G}_i \right|=5$) and high-frequency ($\left| \mathbb{G}_i \right|=50$) trading scenarios. Here, the number of bidders varies from $10$ to $50$. $\left| \mathbb{G}_i \right|$ means the number of winning tasks of bidder $i$, and $\left| \mathbb{G}_i \right|\!+\!2$ is the length of $HashChain_i$. In Fig.~\ref{fig:simu10}, we can observe that SEAL outperforms SAFE in attaining a lower communication overhead especially when the trading frequency is high. The reason is that SEAL integrates the hashchain-based micropayment mechanism to support batch payment to each winning bidder instead of paying at each auction round, thereby alleviating the communication burden in multi-round frequent micropayments.

The auction execution time and communication cost of off-chain auction execution in Intel SGX are evaluated in Fig.~\ref{fig:simu9}. It can be seen that, given the number of bidders ranging from 20 to 120, the auction time in TEE is less than 0.4s and the communication cost in remote attestations between participants and the enclave is less than 250KB, which is efficient for practical deployment.

\textbf{2) Consensus delay.}
Fig.~\ref{fig:simu12} shows the consensus delay for different transaction types defined in Algorithm~\ref{Algorithm1}. Here, for a transaction, its \emph{consensus delay} refers to the latency from being pending to be confirmed in the blockchain. 
As seen in Fig.~\ref{fig:simu12}, the consensus delay is about 0.3$\sim$0.8 seconds for different types of transactions, and it is less than 0.81 seconds for all transaction types, which is efficient for practical offloading services.

\subsection{Economic Efficiencies}\label{subsec:evalution3}

\textbf{1) Average cost of UAV.}
Fig.~\ref{fig:simu2} compares UAV's average cost among {five schemes} under different vehicle densities. Here, the number of tasks is set as $J=200$, and the linear score function $\mathbf{LIN}$ is adopted in BidGuard.
As seen in Fig.~\ref{fig:simu2}, SEAL outperforms {BidGuard, ARMOR, and SAFE} in acquiring a smaller gap with the exhaustive optimal solution. This is because in BidGuard, users need to upload the perturbed combinatorial bids, instead of the real ones, via the exponential mechanism to ensure DP and prevent inference attacks. Thereby, a large auction efficiency decrease can occur in determining optimal winners and payments based on the perturbed bids. {In ARMOR, it consumes considerable computation energy for UAVs in HE operations for bid privacy protection. In SAFE, as analyzed in Table~\ref{tabComComplexity} and Fig.~\ref{fig:simu10}, it involves higher communication cost for UAVs especially under high $\left| \mathbb{G}_i \right|$ when the vehicle density is low.
On contrary, SEAL determines the winners and due payments based on the true bids in plaintext and batch payment method with the help of smart contracts and the trusted processer for frequent micropayments}, resulting in a higher offloading efficiency in the auction.


\begin{figure}[!tbp]\setlength{\abovecaptionskip}{-0.0cm}
\begin{minipage}[t]{0.245\textwidth}
\centering
    \includegraphics[height=3.45cm,width=\linewidth]{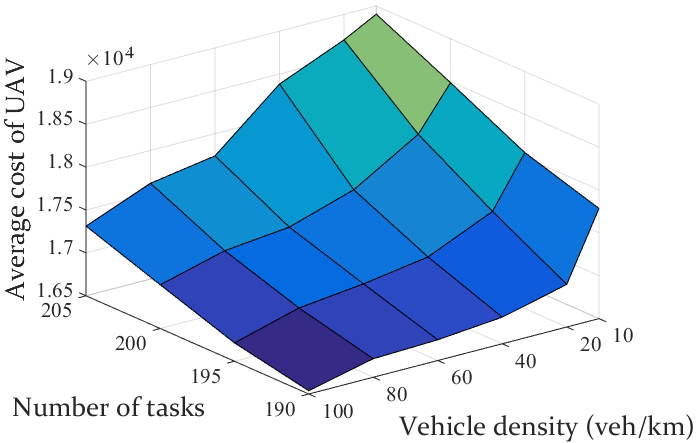}
  \caption{Average cost of UAV vs. vehicle density, with different number of computation tasks.}\label{fig:simu1}
\end{minipage}~~
\begin{minipage}[t]{0.245\textwidth}
\centering
    \includegraphics[height=3.45cm,width=\linewidth]{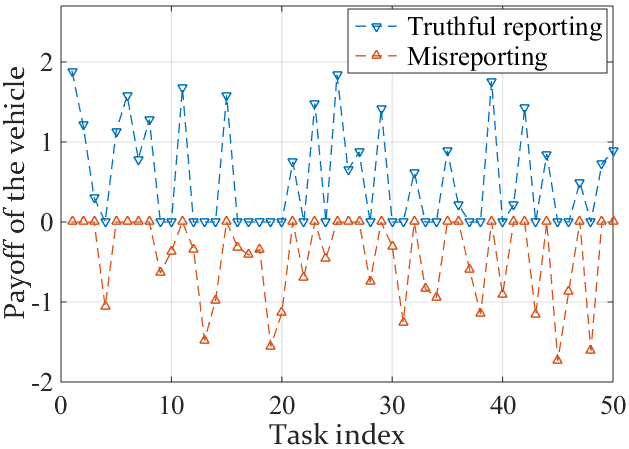}
    \caption{Payoff of a randomly selected vehicle when it submits truthful bids vs. strategic bids in different tasks.}\label{fig:simu5}
\end{minipage}\vspace{-0.25cm}
\end{figure}

Fig.~\ref{fig:simu1} shows the average cost of UAV in its flying route when the vehicle density increases from 10 to 100 veh/km and the number of tasks increases from 190 to 205. It can be seen that the average cost of UAV decreases with the increase of the vehicle density, and it increases with the increase of the number of tasks. The reason is that the UAV can choose vehicles that offer lower costs when the number of candidate vehicles increases. Besides, according to the objective function in Eq. (\ref{eq:optimization}), given the fixed vehicle density, the higher number of tasks can result in a higher cost of the UAV.

\textbf{2) Strategy proofness.}
Fig.~\ref{fig:simu5} shows the payoff of a candidate vehicle in different tasks. It can be seen that the vehicle's payoff under honest participation (i.e., bidding truthfully) is always non-negative and higher than that under strategic bids (i.e., misreporting an arbitrary bid vector), which validates the strategy-proofness of SEAL {in vehicles' bids}.

{\textbf{3) Latency \& energy cost in offloading.} The following conventional offloading} schemes are used for comparison with SEAL in terms of auction efficiency.
\begin{itemize}
  \item \emph{Energy-Aware Auction (EAA) scheme \cite{8761932}:} every task is assigned to the candidate bidder with the minimum energy cost, and the UAV flies at the minimum flying speed $V_{\min}$.
  \item \emph{Delay-Aware Auction (DAA) scheme:} each task is allocated to the bidder with the minimum completion delay, and the UAV's flying speed is fixed at the maximum value. 
  \item \emph{Price-Aware Auction (PAA) scheme:} every task is assigned to the candidate bidder with the minimum bidding price, and UAV's flying speed is randomly selected from $[V_{\min}, V_{\max}]$.
  \item \emph{Cloud-based offloading scheme \cite{8981670}:} the computing tasks of the UAV are offloaded to the remote cloud for processing at each location. Here, $\phi_{\mathrm{cloud}}=8$  and $\chi_{\mathrm{cloud}}^j=10$ GC/s.
  \item \emph{Fog-based offloading scheme \cite{8648099}:} the computing tasks of the UAV are offloaded to the fixed fog server for processing at each location. Here, $\phi_{\mathrm{fog}}=9$  and $\chi_{\mathrm{fog}}^j=3$ GC/s \cite{8672604}. 
\end{itemize}

\begin{figure}[!tbp]\vspace{-0.1cm}
\setlength{\abovecaptionskip}{-0.03cm}
    \subfigure[]{
		\includegraphics[height=3.44cm,width=0.49\linewidth]{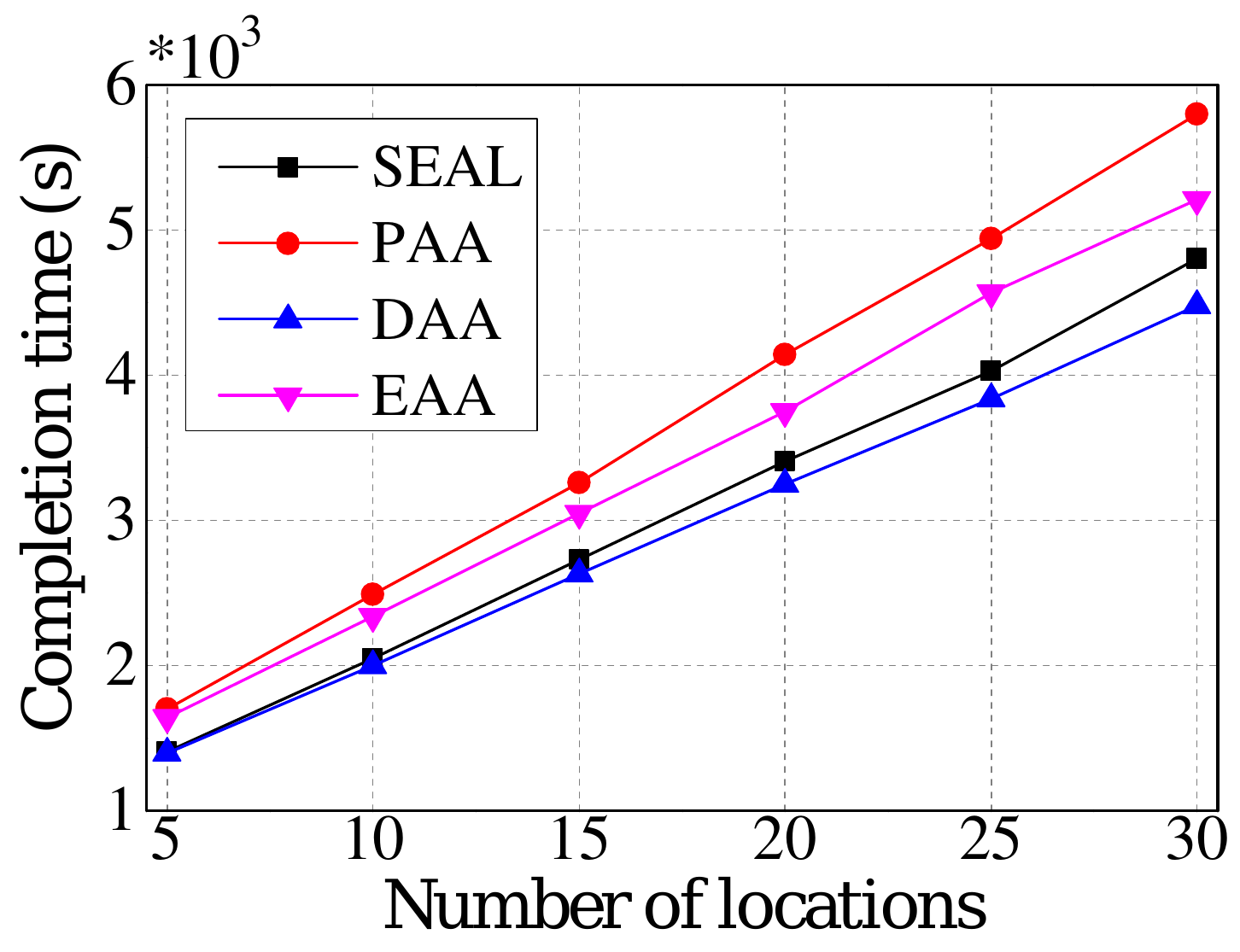}\label{fig:simu3}}
    \subfigure[]{
		\includegraphics[height=3.44cm,width=0.49\linewidth]{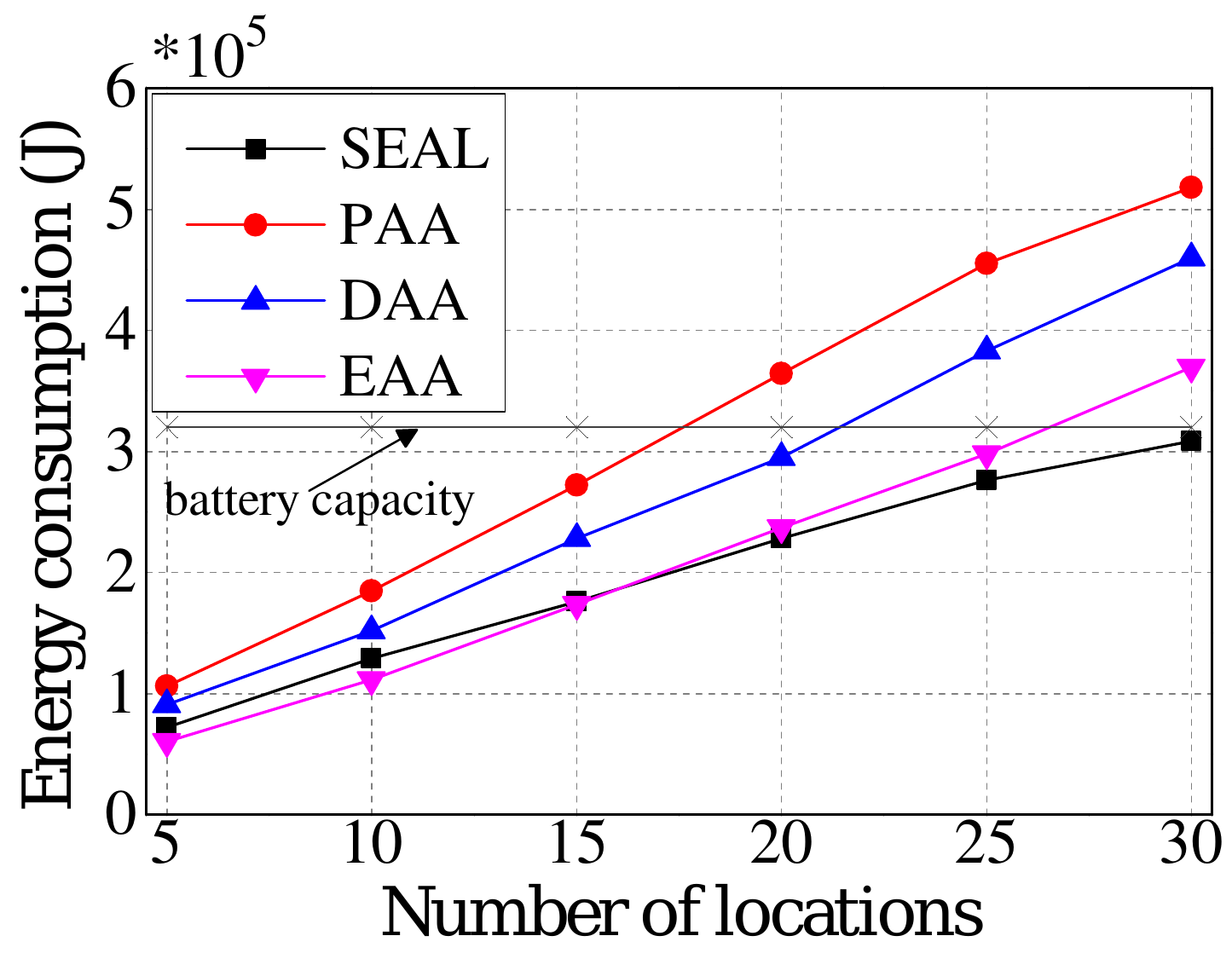}\label{fig:simu4}}
	\caption{Comparison of (a) journey completion time of UAV and (b) energy consumption of UAV vs. number of locations in four schemes.}\label{fig:simu34}
\end{figure}

\begin{figure}[!tbp]\vspace{-0.cm}
\setlength{\abovecaptionskip}{-0.03cm}
    \subfigure[]{
		\includegraphics[height=3.4cm,width=0.492\linewidth]{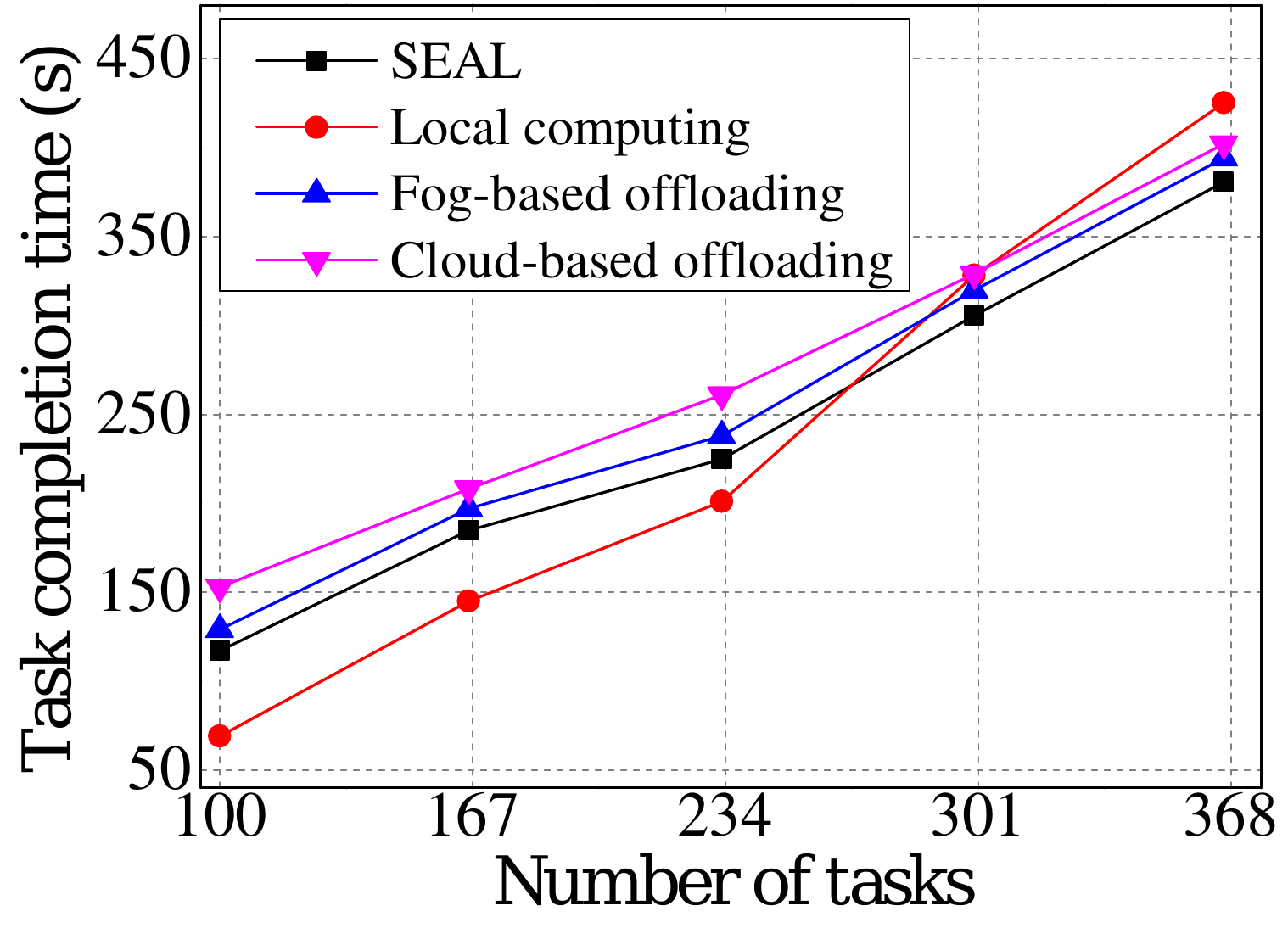}\label{fig:simu7}}
    \subfigure[]{
		\includegraphics[height=3.48cm,width=0.492\linewidth]{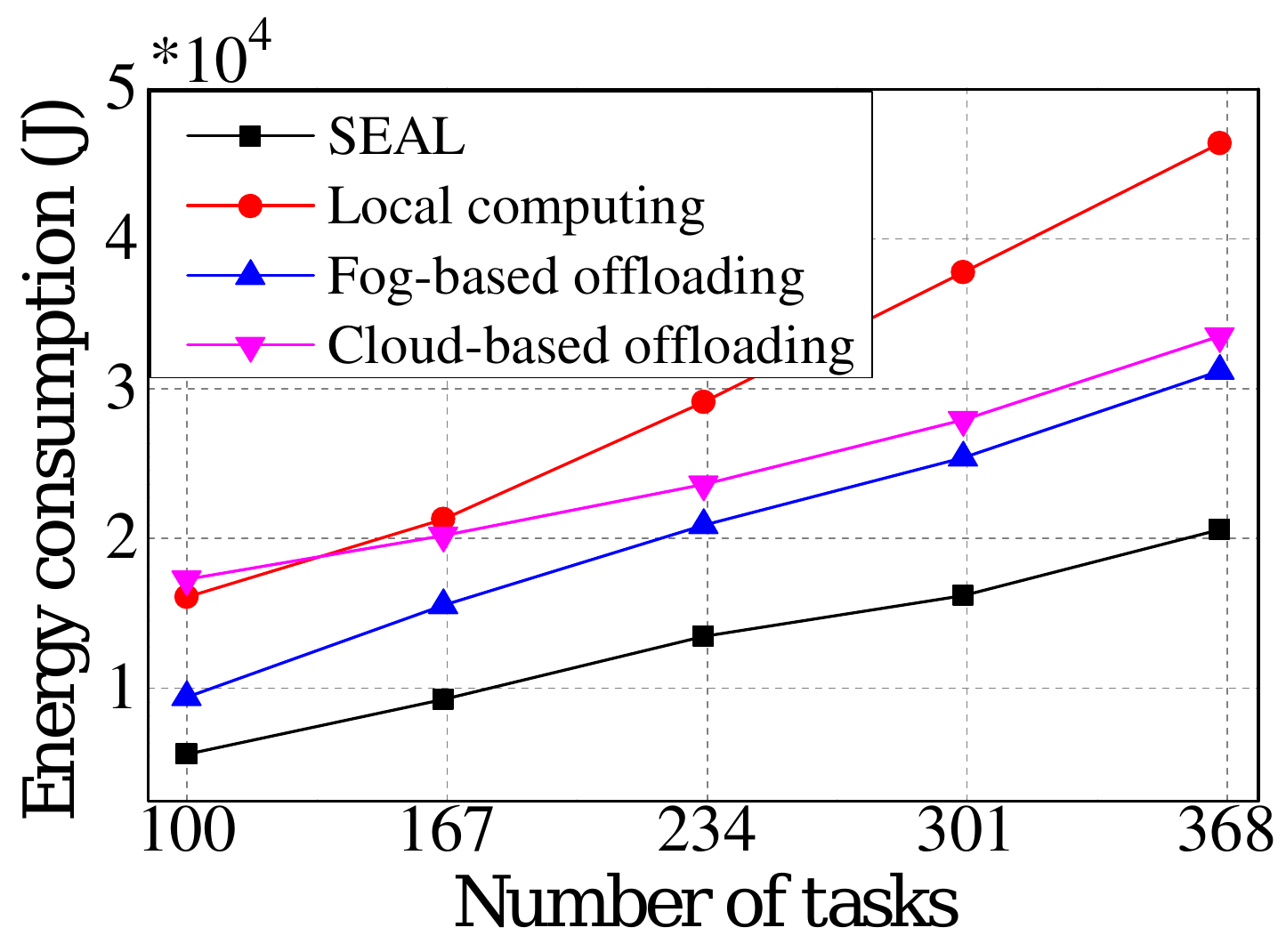}\label{fig:simu8}}
	\caption{Comparison of (a) task completion time and (b) energy consumption of UAV vs. number of tasks in four schemes.}\label{fig:simu78}\vspace{-0.2cm}
\end{figure}

{In the above baselines, as the bid privacy protection and fairness are not considered, these offloading schemes have lower computation and communication overheads than our SEAL scheme. To be objective and fair, we only compare them with our SEAL scheme in terms of auction efficiency (e.g., task completion latency and energy cost) in Figs.~\ref{fig:simu34}-\ref{fig:simu78}.}


Fig.~\ref{fig:simu34} shows the task completion time and energy consumption in four schemes with different number of locations (i.e., $N$).
In Fig.~\ref{fig:simu34}, as more locations cause higher latency in task processing and UAV's transition, both the completion time and energy consumption in four schemes increase with $N$.
In Fig.~\ref{fig:simu3}, DAA attains the smallest task processing delay, and SEAL performs close to DAA when $N$ is small.
In Fig.~\ref{fig:simu4}, when $N$ becomes large, both DAA and PAA incur high energy cost and violate UAV's battery limit. Besides, EAA performs better than SEAL when $N$ is small, while EAA incurs a higher growth rate than SEAL and even violates the battery constraint when $N$ is large. The reason is that a low flying speed raises UAV's energy consumption for lifting against the force of gravity, while a high speed raises the propulsion energy for moving between locations. In PAA, DAA and EAA, UAV flies at the random, maximum, and minimum speed, thereby increasing the energy cost in transitions and violating the battery constraint when $N$ is large.

Fig.~\ref{fig:simu78} shows the task completion time and energy consumption in offloading in four schemes with different number of tasks (i.e., $J$).
In Fig.~\ref{fig:simu7}, the local computing scheme has the smallest completion time when $J$ is small, and it incurs a higher growth rate than other schemes when $J$ becomes large. The reason is that UAV's computation capacity is limited and can be fully occupied by the increasing tasks to be processed. Besides, our SEAL attains the smallest task delay when $J$ is large (in Fig.~\ref{fig:simu7}) and the smallest energy cost in offloading (in Fig.~\ref{fig:simu8}). It is because that vehicles are provisioned with sufficient computing resources and are more close to the UAV for task execution.

\subsection{{Key Insights}}\label{subsec:insights}
\begin{itemize}
  \item {Compared with existing representative privacy-preserving auction approaches based on cryptosystems and TEE, our SEAL achieves lower computation/communication complexity, particularly in high-frequency trading settings. Moreover, our SEAL acquires lower average costs for UAVs than existing representative privacy-preserving auctions. Besides, our SEAL enjoys low off-chain auction execution time within TEE and low consensus delay in the blockchain.}
  \item {Our SEAL ensures fairness, strategy proofness, and privacy preservation simultaneously. Besides, our SEAL enforces high offloading efficiency in terms of low UAV's cost, low task completion delay, and low energy consumption.}
\end{itemize}

\section{Conclusion}\label{sec:CONSLUSION}
In this paper, we have presented SEAL to address efficient, fair, and privacy-preserving computation offloading for UAV applications.
First, we have introduced a VFC-oriented auction-based collaborative mechanism to efficiently offload UAVs' intensive computation missions to ground vehicles while guaranteeing economic robustness.
Then, we have implemented a fair exchange protocol in smart contracts to enforce both participation fairness and exchange fairness between distrustful entities.
By further integrating TEE into smart contracts, an off-chain auction mechanism has been devised to preserve vehicles' privacy in an efficient manner.
At last, simulation results have validated the effectiveness of SEAL in terms of offloading efficiency, cost saving, and system overheads.
For future work, we will further extend SEAL to be resistant to collusive vehicles with bid manipulation prevention.

\begin{appendices}
\section{proof of Theorem 1}\label{Appendix A}
\textbf{Theorem 1}. \emph{A SRC auction mechanism is combinatorial incentive-compatible if the following two properties hold:} 
\begin{itemize}
  \item Monotonicity. \emph{For each task $\Im_{j,n}$, given that other bidders' strategies are fixed, any bidder $i\in \mathbb{I}$ wins the auction with bid $({\chi_{i}^{j}},{b_{i}^{j}})$ still wins by bidding $({{\chi_{i}^{j}}'},{{b_{i}^{j}}'})$  with ${{\chi_{i}^{j}}'}>{\chi_{i}^{j}}$ and ${{b_{i}^{j}}'}<{b_{i}^{j}}$.}
  \item Critical payment. \emph{Any winner $i$ with bid $({\chi_{i}^{j}},{b_{i}^{j}})$ of task $\Im_{j,n}$ is paid the critical payment, i.e., the supremum of all bidding prices ${b_{i}^{j}}$'s such that $({\chi_{i}^{j}},{{b_{i}^{j}}'})$ still wins, i.e., $p_{i*,j}^{\mathrm{CP}}=\mathrm{sup}\{{b_{i}^{j}}' |{\beta_{i}^{j}}'=1\}$, when the bids of others remain unchanged.}
\end{itemize}
\begin{IEEEproof}
Let $(\chi_{i}^{j}, \Theta(\chi_{i}^{j}))$ be the truthful combinatorial bid of vehicle $i$.
Obviously, if vehicle $i\in \mathbb{C}_{j,n}$ loses the auction with untruthful combinatorial bid $(\chi_{i}^{j'}, {b_{i}^{j'}})$ where $\chi_{i}^{j'} \ne \chi_{i}^{j}$ or $b_{i}^{j'} \ne \Theta(\chi_{i}^{j})$, its payoff is non-positive. Besides, rational bidders will not receive negative payoffs. Therefore, only the case that bidder $i$ wins with bid $(\chi_{i}^{j'}, {b_{i}^{j'}})$ needs to be considered.

First, we prove that for a strategic bidder that bids an untruthful computation resource supply $\chi_{i}^{j'} \ne \chi_{i}^{j}$ is always dominated by $\chi_{i}^{j}$. If $\chi_{i}^{j'} > \chi_{i}^{j}$, then even if the bidder wins, it has to supply $\chi_{i}^{j'}$ amount of computation resources. Thereby, the payoff of the vehicle is non-positive.
When the vehicle bids $\chi_{i}^{j'} < \chi_{i}^{j}$, according to the monotonicity property, if it wins with bid $\chi_{i}^{j'}$, it could also win with $\chi_{i}^{j}$. Besides, given the critical payment property, the payment in the latter
case (i.e., $\chi_{i}^{j}$) is never lower than that in the former case (i.e., $\chi_{i}^{j'}$). Therefore, bidding the truthful computation resource supply is the dominant strategy of any vehicle.

Next, we prove that for a strategic bidder that misreports a bidding price $b_{i}^{j'} \ne \Theta(\chi_{i}^{j})$ is always dominated by $\Theta(\chi_{i}^{j})$. If vehicle $i$ wins by bidding $\Theta(\chi_{i}^{j})$ and it is paid the critical value $p_{i,j}^{\mathrm{CP}} > \Theta(\chi_{i}^{j})$, then any possible bidding price $b_{i}^{j'} \leq p_{i,j}^{\mathrm{CP}}$ still leads to the winning. Moreover, under the critical payment condition, it is still paid the same amount $p_{i,j}^{\mathrm{CP}}$, resulting in the same payoff as when bidding $\Theta(\chi_{i}^{j})$. Under this circumstance, a bidding price $b_{i}^{j'} > p_{i,j}^{\mathrm{CP}}$ will lose the auction, and the vehicle obtains the zero payoff at last. Therefore, for the vehicle that wins by bidding $\Theta(\chi_{i}^{j})$, it is not better than truthfully bidding $\Theta(\chi_{i}^{j})$.
If vehicle $i$ loses by bidding $\Theta(\chi_{i}^{j})$, then the largest winning bidding price is $p_{i,j}^{\mathrm{CP}} \leq \Theta(\chi_{i}^{j})$. In this case, bidding at most $p_{i,j}^{\mathrm{CP}}$ will win the auction but yields an negative payoff of the vehicle, and bidding more than $p_{i,j}^{\mathrm{CP}}$ will still lose the auction for the vehicle. Therefore, for the vehicle that loses by bidding $\Theta(\chi_{i}^{j})$, truthfully bidding $\Theta(\chi_{i}^{j})$ is still the dominant strategy.

Thereby, our SRC auction mechanism satisfies CIC if both the monotonicity and critical payment properties hold. 
\end{IEEEproof}

\section{proof of Lemma 1}\label{Appendix B}
\emph{\textbf{Lemma 1}. SEAL satisfies combinatorial incentive compatibility (CIC) for both computation resource supply and bidding price.}
\begin{IEEEproof}
It is equivalent to prove the CIC of each sub-auction $\mathcal{A}_j$ for task $\Im_{j,n}$, $\forall j \in \mathbb{J}_n'$. In other words, we need to prove the truthfulness of each bidder's reported computing resource and the bidding price in each sub-auction $\mathcal{A}_j$.
According to Theorem 1, it suffices to prove the monotonicity of worker selection process in $\mathcal{A}_j$ and payment $p_{i,j}^{\mathrm{CP}}$ is the critical value for the bidder $i$ to win the auction $\mathcal{A}_j$. Without loss of generality, suppose that bidder $i$ wins auction $\mathcal{A}_j$ with the truthful bid $(\chi_{i}^{j},\Theta(\chi_{i}^{j}))$.

\emph{1) Monotonicity.} The monotonicity of $\mathcal{A}_j$ is proved in the following two cases.

Case 1: bidder $i$ decreases its bidding price, i.e., ${b_{i}^{j}}^{-} <\Theta(\chi_{i}^{j})$. The reduction in bidding price decreases the bidder's MCF contribution, i.e.,
\begin{align}
\digamma(\chi _{i}^{j},{b_{i}^{j}}^{-})<\digamma(\chi_{i}^{j},\Theta(\chi_{i}^{j})).
\end{align}
As a result, bidder $i$ is still the winner.

Case 2: bidder $i$ increases its reported amount of computing resource, i.e., ${\chi_{i}^{j}}^{+} \!>\!\chi_{i}^{j}$. This also decreases bidder $i$'s MCF, i.e., \begin{align}
\digamma({\chi_{i}^{j}}^{+},\Theta(\chi_{i}^{j}))<\digamma(\chi_{i}^{j},\Theta(\chi_{i}^{j})).
\end{align}
Consequently, bidder $i$ is still the winner.

According to the above two cases, each sub-auction $\mathcal{A}_j$ is monotone.

\emph{2) Critical payment.} Next, we prove that $p_{i,j}^{\mathrm{CP}}$ derived by Algorithm~2 exactly equals to the critical value. 
The proof is divided into two cases. 

Case 1: given the fixed $\chi_{i}^{j}$, bidder $i$ bids less than or equal to the obtained payment, i.e., ${b_{i}^{j}}^{-} \le p_{i,j}^{\mathrm{CP}}$. Then, bidder $i$ still wins and receives the same payment, as well as the same payoff.

Case 2: given the fixed $\chi_{i}^{j}$, bidder $i$ bids greater than the obtained payment, i.e., ${b_{i}^{j}}^{+} > p_{i,j}^{\mathrm{CP}}$. We assume that bidder $k$ ($k\ne i$) is the new winner (i.e., critical bidder) when bidder $i$ does not participate in auction $\mathcal{A}_j$. In this case, based on Eqs. (27) and (32), we have 
\begin{align}\label{eq:ICprove}
\digamma&(\chi _{i}^{j},{b_{i}^{j}}^{+}) \nonumber\\
&> \varpi s_{j,n}(\frac{P^{\text{hov}}\zeta _{j,n}}{\chi _{i}^{j}}+\frac{P^{\text{A2G}}+P^{\text{hov}}}{\gamma _{i}^{j}}) +(1-\varpi){\lambda _p} p_{i,j}^{\mathrm{CP}} \nonumber \\
&= \varpi s_{j,n}(\frac{P^{\text{hov}}\zeta _{j,n}}{\chi _{k}^j}+\frac{P^{\text{A2G}}+P^{\text{hov}}}{\gamma _{k}^j}) +(1-\varpi){\lambda _p} b_{k}^j \nonumber \\
&= \digamma(\chi _{k}^j,b_{k}^j).
\end{align}
Therefore, bidder $k$ wins the auction while bidder $i$ loses.

Combing the above two cases, when $\chi_{i}^{j}$ is fixed, $p_{i,j}^{\mathrm{CP}}$ is exactly the critical payment for bidder $i$ above which bidder $i$ losses.
It can be concluded that bidding truthfully is the dominant strategy of each bidder to maximize its payoff. Lemma 1 is proved.
\end{IEEEproof}

\section{proof of Lemma 2}\label{Appendix C}
\emph{\textbf{Lemma 2}. SEAL satisfies individual rationality (IR).}
\begin{IEEEproof}
As our SRC auction mechanism satisfies CIC, any bidder will be motivated to submit its truthful bid. Besides, the payoff of any bidder that does not participate in the auction is zero. We consider the following two cases.

Case 1: bidder $i$ wins the auction with the truthful bid $(\chi_{i}^{j},\Theta(\chi_{i}^{j}))$. In this case, suppose that bidder $k\in \mathbb{C}_{j,n}\backslash \{i\}$ is the critical bidder, i.e., virtual winner of task $\Im_{j,n}$.
Similarly, we have $\digamma (\chi _{i}^{j},p_{i,j}^{\mathrm{CP}})=\digamma (\chi _{k}^j,\Theta(\chi_{k}^j))$. Since bidder $k$ will lose if bidder $i$ joins in the auction, we have
\begin{align}
\digamma (\chi _{i}^{j},\Theta(\chi_{i}^{j}))\le \digamma (\chi _{k}^j,\Theta(\chi_{k}^j)) =\digamma (\chi _{i}^{j},p_{i,j}^{\mathrm{CP}}).
\end{align}
According to Eq. (27), we have ${p_{i}^{j}}^{*}\ge \Theta(\chi_{i}^{j})$. Thereby, 
$\pi (\chi _{i}^{j},b_{i}^{j}) = p_{i,j}^{\mathrm{CP}}-\Theta(\chi _{i}^{j}) \ge 0$.

Case 2: bidder $i$ loses the auction with the truthful bid $(\chi_{i}^{j},\Theta(\chi_{i}^{j}))$. In this case, $\pi (\chi _{i}^{j},b_{i}^{j}) = 0$.

Hence, the payoff of any bidder participating in the auction is no less than zero. Lemma 2 is proved.
\end{IEEEproof}

\section{proof of Theorem 3}\label{Appendix C-1}
\emph{\textbf{Theorem 3}. SEAL can preserve participants' bidding privacy.}
\begin{IEEEproof}
{We prove the bidding privacy protection of SEAL in two successive phases: off-chain auction execution and on-chain fair exchange. In the first phase, note} that the private bids involved in the auction execution process are securely processed inside the TEE. Both the UAV and malicious bidders can only observe the encrypted data but no useful bid information from the output of TEE, even if the outside hardware such as I/O or storage is compromised. Therefore, vehicles' private bids $(\chi_{i}^{j},b_{i}^{j})$ in the auction process can be preserved {with the assistance of TEE in the open smart contract systems}.

Next, we analyze the bidding privacy in the fair exchange phase (including Commit, On-chain exchange, Claim, and Refund operations). In this phase, only the winners and their critical payments are made public on blockchain ledgers in the auction outcome.
On one hand, as the payments to winners are recorded in $\mathrm{tx_{commit}}$ in plaintext on transparent blockchain ledgers, adversaries can deduce the private bid information of critical bidders. Nevertheless, since the identities of critical bidders are hidden from the losing bidders in SEAL, their identity privacy can be protected.
On the other hand, in our proposed SRC auction, each winner is paid with the critical payment (calculated based on the critical bidder in Eq. (32)) instead of its raw bid, to ensure combinatorial strategy proofness. Given the winners and their critical payments in the public auction outcome, adversaries cannot link winners' true bidding prices with the corresponding critical payments. As such, the true bidding price of each winner can be hidden and preserved in SEAL. Therefore, vehicles' bid privacy can be preserved in the fair exchange phase.
\end{IEEEproof}

\section{proof of Theorem 4}\label{Appendix C-2}
\emph{\textbf{Theorem 4}. SEAL is a fair auction mechanism, i.e., it guarantees both participation fairness and exchange fairness.}
\begin{IEEEproof}
%
\emph{1) Participation fairness.}
In the blockchain, as each bidder needs to be registered and make sufficient deposits, any registered bidder who aborts the assigned tasks will be punished by confiscating the deposits. As the deposits are greater than any bidding price, any bidder who aborts can obtain a negative payoff regardless of winning or losing the auction. Besides, for an honest bidder, its expected payoff is always larger than that if it aborts. Therefore, any rational and selfish bidder will honestly follow the auction protocol and has no incentives to abort the auction.

\emph{2) Exchange fairness.}
The exchange fairness is guaranteed by the proposed on-chain fair exchange protocol in Algorithm~1.
Specifically, after publishing the metadata information $\{meta_i\}_{i\in \mathbb{W}}$ on blockchain, the hash values in the hashchain (generated by the UAV) can be utilized as commitments for micropayments to bidders to prevent the UAV from refusing to pay.
Besides, after the off-chain delivery of hash values, if the bidder refuses to release task results to the UAV or delivers an incorrect key for decryption, its misbehavior will be immutably recorded in the blockchain and accordingly the smart contract will not conduct the corresponding payment to the bidder from the escrow pool.
Under the supervision of smart contracts, the on-chain exchange process of payment and task results can be executed automatically and atomically between distrustful bidders and the UAV.

Hence, both participation fairness and exchange fairness are ensured in SEAL. 
\end{IEEEproof}

\section{proof of Theorem 5}\label{Appendix D}
\emph{\textbf{Theorem 5}. The computational complexity and communication complexity of SEAL are $\mathcal{O}(J_n I \log(I))$ and $\mathcal{O}(I (M \cdot bit_{\pi}+ bit_{c}))$, respectively.}
\begin{IEEEproof}
Let $bit_p$, $bit_c$, and $bit_{\pi}$ be the bit lengths of the plaintext of combinatorial bid, ciphertext and signature (we consider them to be the same by default), and ZKP, respectively.
Let $M$ be the number of consensus nodes in the smart contract system.

In the \textbf{init} phase, the main computation is to encrypt the bids, resulting in a complexity of $\mathcal{O}(I)$. ${I}$ is the total number of vehicles. The submission of ciphertext of bids to the TEE costs $\mathcal{O}(I\cdot bit_c)$.
In the \textbf{off-chain auction execution} phase, the computation and communication complexities are greatly optimized due to the computing over the plaintext inside the TEE.
In this phase, there exist two main operations for winner determination and pricing, i.e., bid sorting and comparison, where the total computational complexity yields $\mathcal{O}(\sum\nolimits_{j=1}^{J_n}{I_j} + J_n\left|\mathbb{C}_{j,n}\right|\log(\left|\mathbb{C}_{j,n}\right|))$. ${I_j}$ is the number of vehicles involved in task $\Im_{j,n}$. After the off-chain auction execution, publishing auction result $(\overrightarrow{\beta},\overrightarrow{p})$ incurs a $\mathcal{O}(M\cdot bit_p)$ communication complexity.

In the \textbf{commit} phase, the computation overhead mainly due to the generation of hashchains for winners, which yields $\mathcal{O}(\sum\nolimits_{i=1}^{\left|\mathbb{W}\right|}{\left|\mathbb{G}_{i}\right|})$. The communication only occurs in sending the transaction $\mathrm{tx_{commit}}$ to the smart contract by the UAV.
In the \textbf{on-chain exchange} phase, the computation overhead mainly consists of the encryption/decryption of processed results and symmetric key, as well as the generation and verification of ZKP, which yields $\mathcal{O}(\sum\nolimits_{i=1}^{\left|\mathbb{W}\right|}{\left|\mathbb{G}_{i}\right|})$. For exchange of task results and payments, each winner delivers the task result message $ResMsg$ and decryption key $k_i^l$ to the UAV and receives the hash values as micropayments from the UAV, which yields $\mathcal{O}(M\left|\mathbb{W}\right|\cdot bit_{\pi})$.
In the \textbf{claim} phase, each winner computes its due payment, which incurs a $\mathcal{O}(\sum\nolimits_{i=1}^{\left|\mathbb{W}\right|}{\left|\mathbb{G}_{i}\right|})$ computation overhead as $N\le \left|\mathbb{G}_{i}\right|$. Meanwhile, the communication complexity is $\mathcal{O}(M\left|\mathbb{W}\right|)$.
It is worth mentioning that in the on-chain exchange and claim phases, all winners perform task result/payment exchange and redeem the due payment in parallel. Thereby, the system latency can be further alleviated.

In both \textbf{deposit} and \textbf{refund} phases, the computation and communication overheads yield $\mathcal{O}(I)$ and $\mathcal{O}(I M)$ in financial settlement, respectively.

Therefore, the overall computational complexity of SEAL is $\mathcal{O}(\sum\nolimits_{j=1}^{J_n}{I_j} + J_n\left|\mathbb{C}_{j,n}\right|\log(\left|\mathbb{C}_{j,n}\right|) + \sum\nolimits_{i=1}^{\left|\mathbb{W}\right|}{\left|\mathbb{G}_{i}\right|})$, which can be simplified as $\mathcal{O}(J_n I \log(I))$. And it determines winners and payments in polynomial time. Moreover, the overall communication complexity of SEAL is $\mathcal{O}(I M bit_{\pi}+I bit_{c})$. 
\end{IEEEproof}

\end{appendices}

\bibliographystyle{IEEETran}
\bibliography{New_ref}
\begin{IEEEbiography}[{\includegraphics[width=1in,height=1.25in,clip,keepaspectratio]{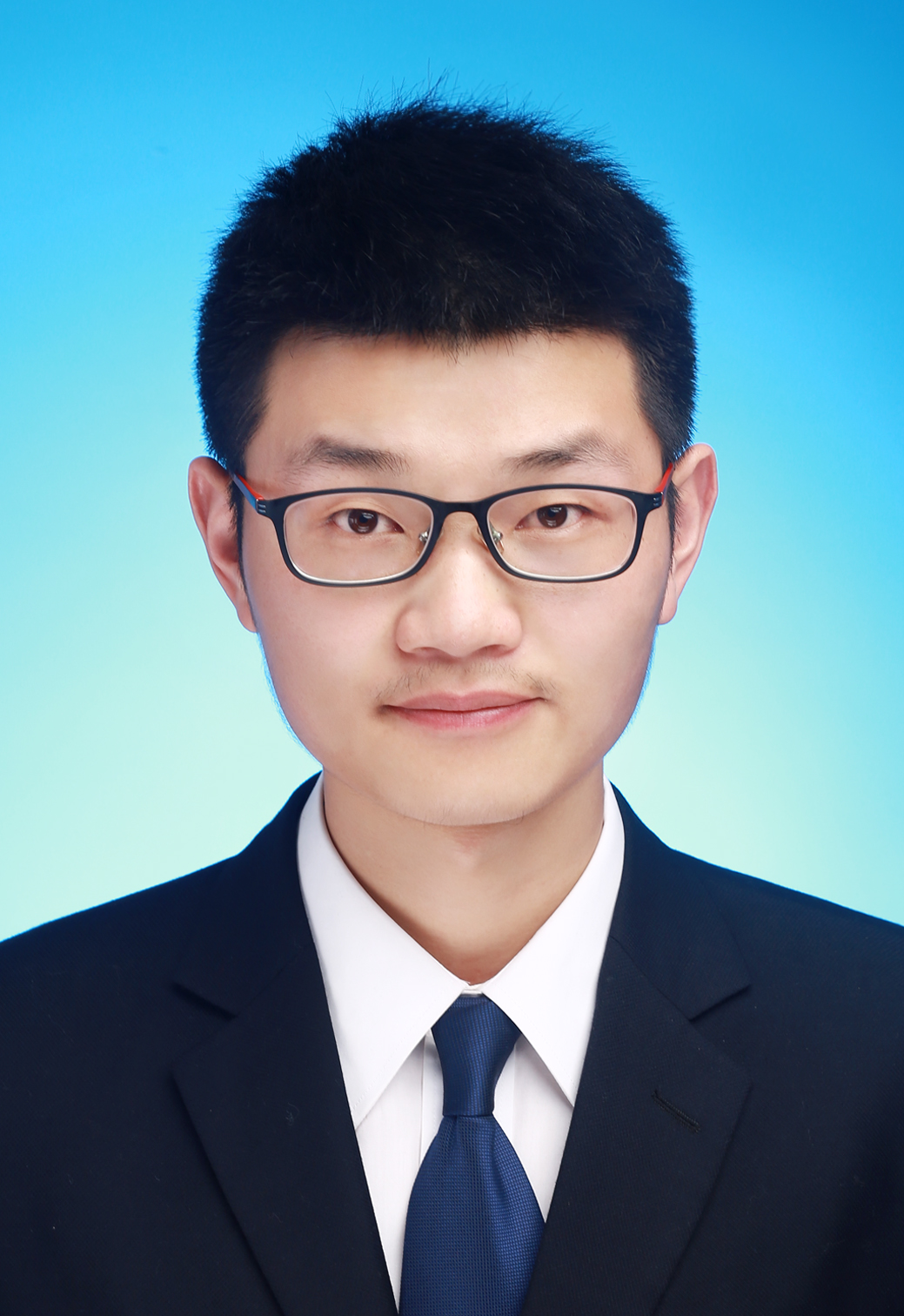}}]{Yuntao Wang}
received the Ph.D degree in Cyberspace Security from Xi'an Jiaotong University, Xi'an, China, in 2022, where he is currently an Assistant Professor with the School of Cyber Science and Engineering. His research interests include security and privacy in intelligent IoT, network games, and blockchain.
\end{IEEEbiography}\vspace{-1cm}

\begin{IEEEbiography}[{\includegraphics[width=1in,height=1.25in,clip,keepaspectratio]{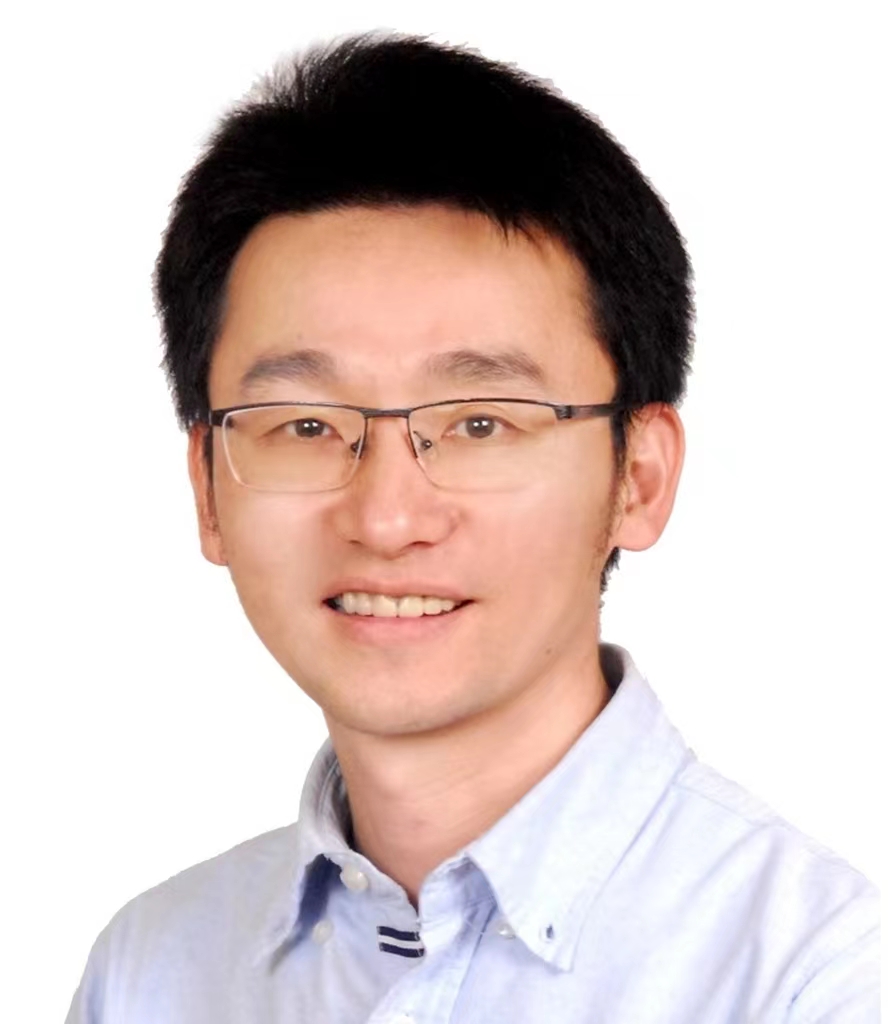}}]{Zhou Su}
has published technical papers, including top journals and top conferences, such as {\scshape IEEE Journal on Selected Areas in Communications}, {\scshape IEEE Transactions on Information Forensics and Security}, {\scshape IEEE Transactions on Dependable and Secure Computing}, {\scshape IEEE Transactions on Mobile Computing}, {\scshape IEEE/ACM Transactions on Networking}, and {\scshape INFOCOM}. 
Dr. Su received the Best Paper Award of International Conference IEEE ICC2020, IEEE BigdataSE2019, and IEEE CyberSciTech2017. He is an Associate Editor of {\scshape IEEE Internet of Things Journal}, {\scshape IEEE Open Journal of the Computer Society}, and {\scshape IET Communications}.
\end{IEEEbiography}\vspace{-1cm}

\begin{IEEEbiography}[{\includegraphics[width=1in,height=1.25in,clip,keepaspectratio]{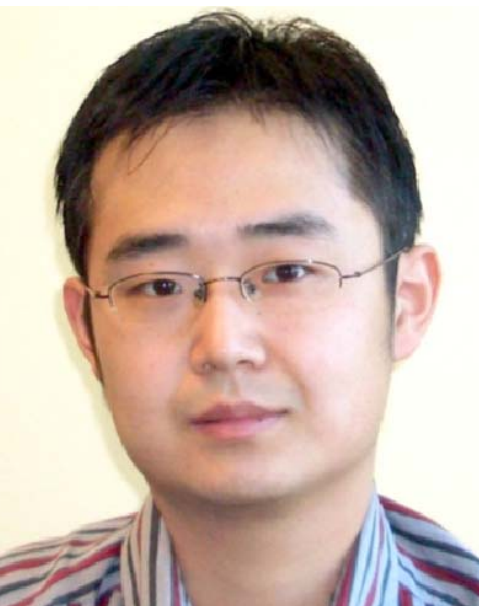}}]{Tom H. Luan}
received the Ph.D. degree from the University of Waterloo, Canada, in 2012. He is currently a Professor with Xi'an Jiaotong University, China. He has authored/coauthored more than 97 journal articles and 58 technical articles in conference proceedings. His research mainly focuses on content distribution and media streaming in vehicular ad hoc networks and peer-to-peer networking and the protocol design and performance evaluation of wireless cloud computing and edge computing.
\end{IEEEbiography}\vspace{-1cm}

\begin{IEEEbiography}[{\includegraphics[width=1in,height=1.25in,clip,keepaspectratio]{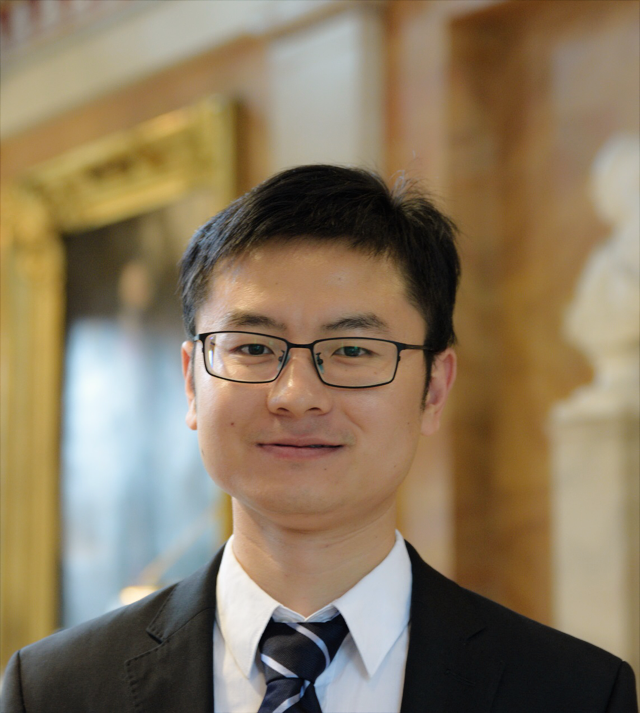}}]{Jiliang Li}
received the Dr. rer. nat. degree in computer science from the University of G\"{o}ettingen, G\"{o}ettingen, Germany, in 2019. He is currently a Researcher Professor and PhD Supervisor with the School of Cyber Science and Engineering, Xi'an Jiaotong University, Xi'an, China. His research interests include information security, cryptography, blockchain and IoT security.
\end{IEEEbiography}\vspace{-1cm}

\begin{IEEEbiography}[{\includegraphics[width=1in,height=1.25in,clip,keepaspectratio]{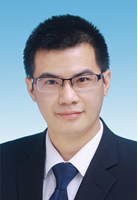}}]{Qichao Xu}
received the Ph.D degree from the school of Mechatronic Engineering and Automation, Shanghai University, Shanghai, China, in 2019. He is currently an Associate Professor with Shanghai university. His research interests are in trust and security, the general area of wireless network architecture, Internet of things, vehicular networks, and resource allocation. He has published more than 50 papers in some respected journals, e.g., IEEE TIFS, IEEE TDSC, IEEE TWC, IEEE TII, IEEE TVT, etc. He was receipt of the best paper awards from several international conferences including IEEE IWCMC2022, IEEE MSN2020, EAI MONAMI2020, IEEE Comsoc GCCTC2018, IEEE CyberSciTech 2017, and WiCon2016.
\end{IEEEbiography}\vspace{-1cm}

\begin{IEEEbiography}[{\includegraphics[width=1in,height=1.25in,clip,keepaspectratio]{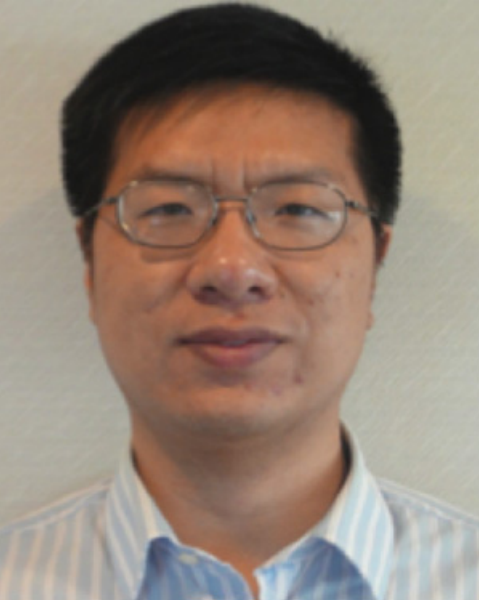}}]{Ruidong Li}
received the D.Eng. degree from the University of Tsukuba in 2008. He is currently an Associate Professor with the College of Science and Engineering, Kanazawa University, Japan. His research interests include future networks, big data networking, blockchain, and network security. He is the Secretary of IEEE ComSoC Internet Technical Committee and the Founder and Chair of the IEEE SIG on big data intelligent networking and IEEE SIG on intelligent Internet edge.
He is a guest editor of prestigious journals, such as {\scshape IEEE Communications Magazine}, {\scshape IEEE Network Magazine}, and {\scshape IEEE Transactions on Network Science and Engineering}.
\end{IEEEbiography}

\end{document}